\documentclass{article}

\usepackage[utf8]{inputenc} 
\usepackage[T1]{fontenc}    
\usepackage{hyperref}       
\usepackage{url}            
\usepackage{booktabs}       
\usepackage{amsfonts}       
\usepackage{nicefrac}       
\usepackage{microtype}      
\usepackage{lipsum}		
\usepackage{graphicx}
\usepackage{natbib}
\usepackage{doi}
\RequirePackage{amsthm,amsmath,amsfonts,amssymb}
\hypersetup{colorlinks=true, linkcolor=blue, citecolor=blue, urlcolor=blue}

\title{Likelihood-Based Confidence Curves for Bounded Parameters}

\title{Likelihood-Based Confidence Curves for Bounded Parameters}

\author{%
  Hedvig Borgen Reiersrud, Clara Bertinelli Salucci,\\
  Riccardo De Bin, Alessandra Rosalba Brazzale
}

\date{October 2026}   

\begin{document}
\maketitle

\begin{abstract}
When a parameter is bounded, the likelihood ratio statistic does not generally follow the chi-square limit distribution of Wilks' theorem. In fact, its asymptotic null distribution depends on the distance from the null value to the boundary, interpolating between the chi-bar-square law on the boundary and the chi-square law in the deep interior of the parameter space. We derive this distribution and use it to construct confidence curves, which are nested collections of two-sided confidence intervals or regions, for bounded parameters. We extend the existing result for the scalar parameter to a two-dimensional parameter with one or both components bounded. As the bivariate distributions are integrals of normal distributions, which are not reducible to standard statistical functions, we also give a simpler approximated distribution obtained by moment-matching. We illustrate the methods on a canonical analysis from high-energy physics, namely the Poisson signal-plus-background experiment.\end{abstract}


\section{Introduction}
\label{sec:intro}
Classical inference often depends on Wilks' theorem \citep{Wilks1938}: under standard regularity conditions the likelihood ratio statistic has an asymptotic chi-square null distribution, providing the reference distribution for tests and confidence intervals. One of these regularity conditions is not met when the hypothesised null value of the parameter lies on the boundary of the parameter space \citep[see, e.g.,][]{BrazzaleMameli2024}. For the simple case of testing a one-dimensional parameter on the boundary, \citet{Chernoff1954} showed that the limiting law is a mixture of a point mass in zero and a chi-square with one degree of freedom. This work was later extended to the multivariate case mainly by \cite{shapiro_asymptotic_1985, shapiro_towards_1988}, \citet{SelfLiang1987} and \citet{kopylev_asymptotic_2011}. Together, these results form a well-developed theory for testing \emph{on} the boundary, that is, when the parameter value specified under the null hypothesis lies exactly on the boundary of the parameter space. In this setting, provided there is no nuisance parameter constrained at the boundary, the null distribution is corrected from a chi-square to a chi-bar-square \citep{kudo_multivariate_1963}, and fixed-level tests are adjusted accordingly.

Confidence intervals, however, require more than the distribution at a single boundary point. A confidence set is obtained by inverting a family of tests across a range of null values. As the null candidate approaches the boundary from the inside of the parameter space, the distribution of the likelihood ratio statistic passes from the standard chi-square law to the boundary chi-bar-square through a certain transition regime. Accurate intervals therefore need the null distribution not only on the boundary, where the chi-bar-square correction applies, but throughout all of its neighbourhood. In this article, we study this near-boundary distribution.

A canonical example from high-energy physics involving confidence intervals for bounded parameters is the signal-plus-background experiment: the search for a small signal against a background, a setting discussed in the debate prompted by \citet{Mandelkern2002}. In this analysis, a detector records a count that is assumed to be Poisson-distributed with mean $\mu + \beta$, where $\beta > 0$ is the background rate expected from pre-established physical phenomena and $\mu \ge 0$ the signal rate of potential new physics. The task is to summarise what the data say about $\mu$---in particular, how large it could plausibly be---through an appropriate confidence interval. 

There are two main likelihood-based approaches to this problem. The first ignores the boundary at the inference stage: the likelihood ratio statistic is built from the unrestricted maximiser $\tilde\mu$, which may take values outside the parameter space of $\mu$ provided that the likelihood is mathematically defined there, and the statistic retains its usual chi-square limiting distribution. The boundary constraint $\mu \geq0$ is imposed post hoc, by truncating the resulting intervals to the parameter space at the reporting stage; see, for instance, \citet{feng_statistical_1992}. Throughout this paper, we refer to this method as the \emph{naïve} approach, and $\tilde\mu$ as the naïve estimator. In the context of the signal-plus-background experiment, this approach is discussed in \cite{Mandelkern2002}, and is also used in, for instance, \citet{FraserReidWong2004} and in the experimental papers \citet{babar_collaboration_search_2006} and \citet{ablikim_search_2026}. The second approach works directly with the restricted parameter space, and defines the likelihood ratio statistic in terms of the proper maximum likelihood estimator (MLE), which, by definition, can only take values inside the parameter space. This approach is considered in \citet{FeldmanCousins1998} and \citet{cowan_asymptotic_2011}, and used in several analyses in high-energy physics, for instance in \cite{aaboud_search_2017, aad_search_2023, aad_search_2024}, \cite{abratenko_search_2025} and \cite{nova_collaboration_dual-baseline_2025}.

\citet{Mandelkern2002} observed that the naïve approach can return empty intervals at nominal levels as high as $68\%$ or even $95\%$, which is an outcome that is unsatisfactory to report. In response, \citet{FraserReidWong2004} argued against reducing the inferential information to a single fixed-level confidence interval; such an interval treats inference as a decision, accepting the parameter values that fall inside and rejecting those outside, rather than conveying how much evidence the data carry for each. They proposed reporting the entire $p$-value function instead.  We follow \citet{SchwederHjort2002} and \citet{XieSingh2013} in calling this $p$-value function a \emph{confidence distribution}, which encodes one-sided confidence statements at every parameter value; its two-sided form, the confidence curve, carries the whole nested family of two-sided intervals at all levels in a single object.

Here, we build confidence curves from likelihood ratio statistics based on the MLE, combining the confidence distribution viewpoint of \citet{FraserReidWong2004} with the boundary-respecting approach of \citet{FeldmanCousins1998}. Since a confidence curve, like a confidence interval, is obtained by inverting a test at every candidate value $\theta \in \Theta$, the null distribution of the test statistic must be known (or approximated) at each such $\theta$. \citet{FeldmanCousins1998} obtain the required null distribution pointwise and numerically, through numerical integration for a one-dimensional parameter of interest or by Monte Carlo (MC) simulation for a multidimensional one. \citet[Section 4.2]{SchwederHjort2016} likewise obtain the null distribution by simulation. \citet{cowan_asymptotic_2011} and \citet{WuNeale2012} instead derive it analytically, as a piecewise function of standard normal cumulative distribution functions (CDFs), for a single bounded parameter of interest. This distribution interpolates between the chi-bar-square on the boundary and the ordinary chi-square in the interior, with the transition governed by the standardised distance from the null value to the boundary.

In this paper, we extend the scalar results of, for instance, \citet{WuNeale2012} to a two-dimensional parameter of interest: the mean vector of a bivariate normal distribution with one or both components bounded. In both cases we derive the exact null distribution at any arbitrary null value, recovering the chi-bar-square distributions of \citet{SelfLiang1987} on the boundary, and the standard chi-square in the deep interior. To our knowledge, these bivariate results are new.

In general these distributions have no closed form in terms of standard statistical functions, but they reduce to integrals of normal and chi-square distribution functions that can be evaluated numerically. For practical use, we also give a moment-matched approximation that requires only standard statistical functions. The approximation is exact on the boundary and in the deep interior, and accurate at intermediate null values.

These null distributions are derived analytically for a Gaussian model with known covariance. This case is of direct applied interest in its own right, and it is also the large-sample limit of regular parametric models. As the sample size diverges, however, the standardised distance to the boundary grows without bound for any interior null value, so close-to-boundary effects vanish in the asymptotic limit. In practice, one typically operates in a medium-sample regime, where normality already holds to good approximation but close-to-boundary effects are still present; see \citet{WuNeale2012} for details. It is precisely in this regime where the corrections matter, and neglecting them can lead to severe miscoverage.

The rest of the paper consists of the following: Section~\ref{sec:cc} introduces confidence distributions and confidence curves and their test inversion construction from the likelihood ratio statistic. Section~\ref{sec:dist} derives the boundary corrections for normally distributed data, treating first the bounded scalar mean and later the extension to the two-dimensional mean with one and with both components bounded. Section~\ref{sec:calibration} presents the moment-matched approximation to the distributions from the two-parameter case, while Section~\ref{sec:numerical-evaluation} evaluates the coverage and power of the confidence curves for bounded parameters. Section~\ref{sec:application} applies the methods to the signal-plus-background counting experiment, and Section~\ref{sec:discussion} concludes.

\section{Confidence curves from the  likelihood ratio }\label{sec:cc}
A confidence distribution, $C(\theta; X)$, which is also often referred to as a $p$-value function or significance function, is a sample-dependent distribution estimator of the parameter of interest $\theta \in \Theta$, with fixed but unknown true value $\theta_0$. For fixed $X=x$, it is a cumulative distribution function in $\theta$, representing one-sided confidence statements directly. That is, each possible value of $\theta$ serves as the null value, and $C(\theta; x)$ is the corresponding $p$-value for testing whether the true parameter is at most $\theta$. Thus, a single curve carries the outcome of every such test at once. The confidence distribution is termed \emph{exact} when all of its confidence intervals attain their nominal coverage, or, equivalently, when $C(\theta_0; X) \sim U(0,1)$. For a formal definition of confidence distributions, see \citet[Definition~1]{XieSingh2013}. 

\begin{figure}
    \centering
    \includegraphics[width=0.49\linewidth]{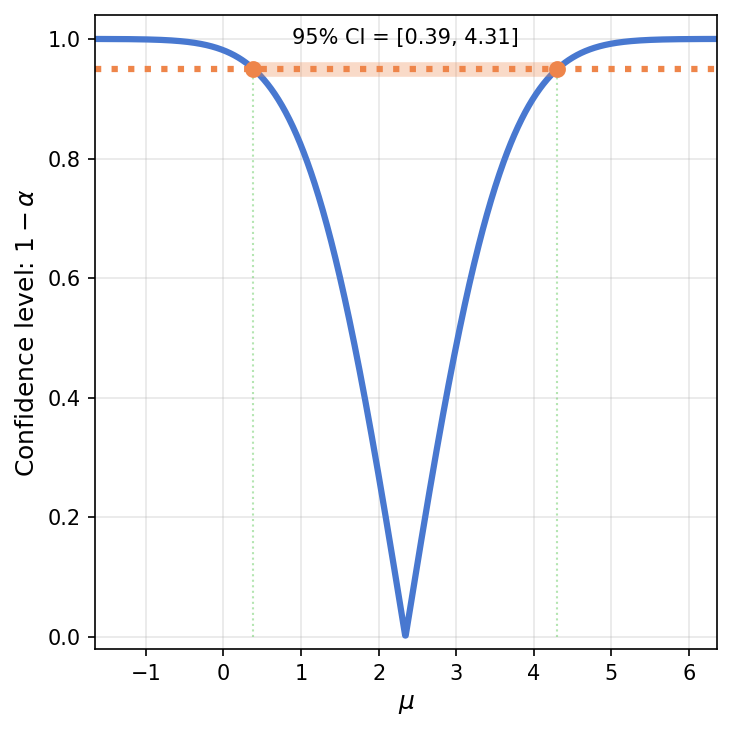}
    \includegraphics[width=0.49\linewidth]{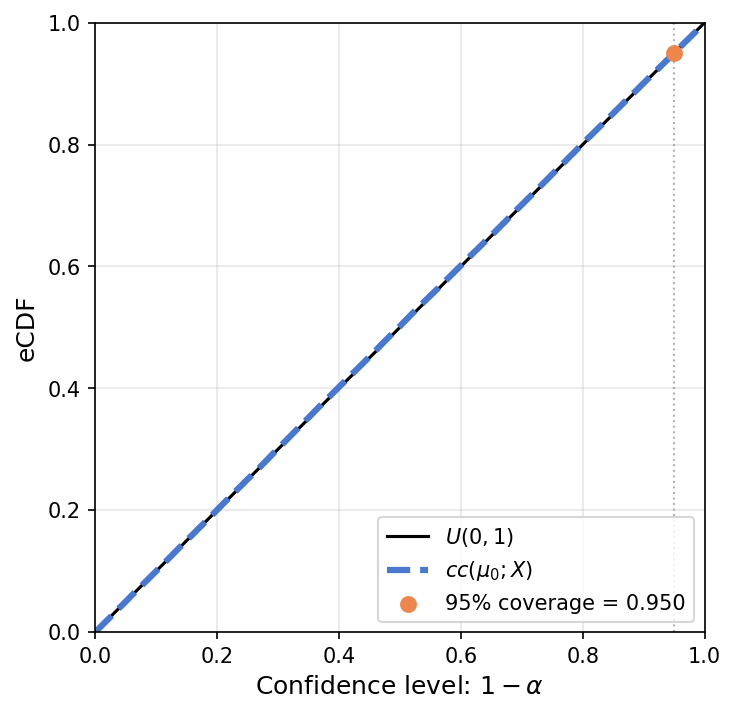}
    \caption{Left: an example confidence curve from one observation $X=x$, where $X\sim\mathcal{N}(\mu_0, 1)$. The horizontal dotted line gives the 95\% CI. Right: the simulated CDF of $cc(\mu_0; X)$ compared to the $U(0,1)$ distribution based on $10^5$ MC simulations.}
    \label{fig:cc_example}
\end{figure}

Throughout this paper, we work with its two-sided counterpart, the confidence curve, which is based on the corresponding two-sided tests. In the univariate setting the two are related by
\begin{equation}
    cc(\theta; X) = \left| 1 - 2C(\theta; X) \right|.
    \label{eq:CDtoCC-transformation}
\end{equation}
An exact confidence curve also satisfies the uniformity condition: $cc(\theta_0;X) \sim U(0,1)$, as the relation in Eq.~\eqref{eq:CDtoCC-transformation} preserves uniformity, so that
$$\operatorname{Pr}_{\theta_0}\left\{cc(\theta_0; X) \leq 1-\alpha\right\} = 1-\alpha, \quad \forall\alpha \in [0,1].$$

Both the confidence distribution and the confidence curve are constructed by test inversion. Given a known analytic pivot, they are obtained by transforming the pivot through its CDF \citep[Section~2.3]{XieSingh2013}. The construction considered here uses the likelihood ratio statistic as an asymptotic pivot. For a log-likelihood $\ell(\theta; x)$, the likelihood ratio statistic is
\begin{equation*}
    q(\theta; x) = 2\{\ell(\hat{\theta}; x) - \ell(\theta; x)\},
\end{equation*}
with $\hat{\theta}$ being the MLE. Under the regularity conditions of \citet{Wilks1938}, $q(\theta_0; X)$ converges in distribution to a $\chi^2_k$, where $k = \dim\theta$, giving the asymptotic confidence curve
\begin{equation}
    cc(\theta; X) = F_{\chi^2_k}\{q(\theta; X)\}.
    \label{eq:cc_LR}
\end{equation}
This construction is discussed in detail in \citet[Chapter~3]{SchwederHjort2016} and used, for instance, by \citet{CunenHjort2022}.

The confidence distributions of \citet{FraserReidWong2004}  and \citet{DavisonSartori2008} are instead based on the signed likelihood root $$r(\theta; X) = \operatorname{sgn}(\hat{\theta} - \theta)\sqrt{q(\theta; X)},$$ which is asymptotically standard normal under the correctly specified model, so that $C(\theta_0; X) = \Phi\{r(\theta_0; X)\}$ is asymptotically $U(0,1)$. The sign supposes an ordering of the parameter space. For a parameter of dimension $k \ge 2$ no such ordering exists and the confidence distribution is therefore not generally well defined in higher dimensions. The confidence curve, however, which depends only on the unsigned likelihood ratio statistic $q(\theta; X) = r^2(\theta; X)$, extends naturally to any dimension; see \citet[Chapter~9]{SchwederHjort2016} for a complete treatment of multivariate confidence functions.

For normal data with known variance, for instance $X \sim \mathcal{N}(\mu, 1)$, the likelihood ratio statistic is $q(\mu; X) = (X - \mu)^2$ which is exactly $\chi^2_1$ under the true value $\mu_0$, and the confidence curve $cc(\mu; X) = F_{\chi^2_1}\{(X - \mu)^2\}$ is exact: $cc(\mu_0; X) \sim U(0,1)$. The left panel of Figure~\ref{fig:cc_example} displays such a curve, with the horizontal dotted line marking the $95\%$ interval; the right panel shows the empirical CDF of $cc(\mu_0; X)$ coinciding with the $U(0,1)$ diagonal, confirming exactness.

For non-normal data, the confidence curve based on Wilks' theorem is only asymptotically exact, and is therefore termed an asymptotic confidence curve. Higher-order corrections, such as the modified likelihood root \citep{barndorff-nielsen_modified_1991}, improve its finite-sample accuracy, as discussed in \citet[Chapter~7]{SchwederHjort2016}. When the parameter is bounded, however, the likelihood ratio statistic is generally not even asymptotically chi-square, and the resulting confidence curve can severely miscover.

\section{Boundary-corrected likelihood ratio distributions}\label{sec:dist}
\begin{figure}
    \centering
    \includegraphics[width=0.49\linewidth]{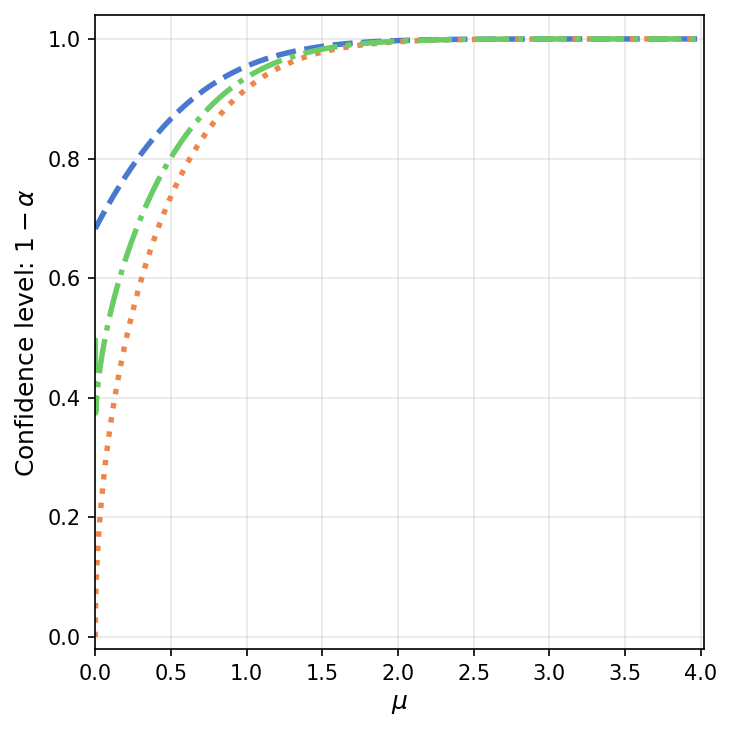}
    \includegraphics[width=0.49\linewidth]{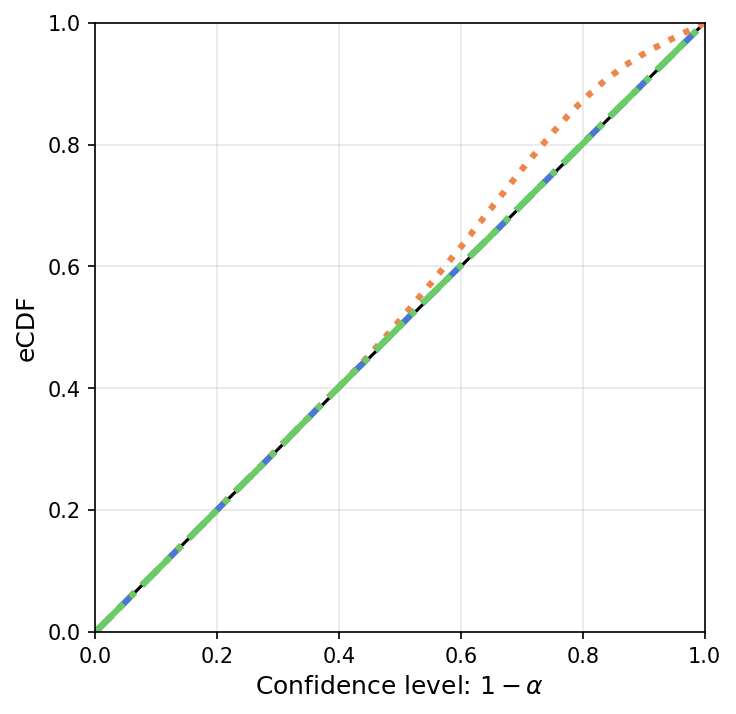}

    \includegraphics[width=0.8\linewidth]{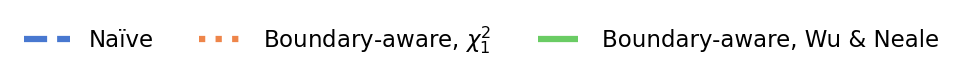}
    \caption{Left: example confidence curves from three different constructions from one observation $x = -1$ stemming from $X\sim\mathcal{N}(\mu, 1)$, where $\mu \in [0, \infty)$. Right: the simulated CDFs of the three $cc(\mu_0; X)$, where $\mu_0 = 0.5$, compared to the nominal $U(0,1)$ distribution based on $10^5$ MC simulations.}
    \label{fig:cc_constrained_example}
\end{figure}

\subsection{Scalar parameter case}\label{sec:1d-WuNeale}
Consider a random sample $X \sim \mathcal{N}(\mu, \sigma^2)$, with unknown true mean $\mu_0$ and known variance $\sigma^2$. Let $M_1 = [0, \infty)$ be the parameter space for $\mu$. The naïve MLE $\tilde{\mu}(X)$ maximises the likelihood over the space where it is mathematically defined, here $\tilde{\mu}(X) = X$. The proper MLE, however, can only take values inside $M_1$,
\begin{equation*}
    \hat{\mu}(X) = \operatorname{argmin}_{\mu \in M_1} (X - \mu)^2 = \max\{0, \tilde{\mu}(X)\}.
\end{equation*}
The two estimators differ only when $\tilde{\mu}(X) < 0$, and the probability of this occurring increases as the null value $\mu$ approaches the boundary. On the boundary, $\mu = 0$, it happens with probability 0.5. The likelihood ratio statistic is then zero, which gives a point mass at zero, conventionally denoted by $\chi^2_0$. With the remaining probability 0.5, $\tilde{\mu}(X)$ lies in the interior and the usual $\chi^2_1$ contribution is recovered. Under the boundary null hypothesis, the statistic therefore follows the chi-bar-square distribution $0.5\chi^2_0 + 0.5\chi^2_1$ \citep{Chernoff1954}. For a general $\mu$, \citet{WuNeale2012} derive the exact CDF of the likelihood ratio statistic as a piecewise function of the observed statistic $t = q(\mu; x)$,
\begin{equation}
    F(t) = \begin{cases}
        F_{\chi^2_1}(t), & \text{if } 0 <t < \delta^2, \\[1ex]
        \Phi(\sqrt{t}) - \Phi\left(-\frac{t+\delta^2}{2\delta}\right), & \text{if } t \geq \delta^2,
    \end{cases}
    \label{eq:CDF-univariate}
\end{equation}
where $\delta = \mu/\sigma$ is the standardised distance from the null value to the boundary, and $\Phi$ and $F_{\chi^2_1}$ are the CDFs of the standard normal and $\chi^2_1$ distributions, respectively. The same distribution appears in \citet[Eq.~(44)]{cowan_asymptotic_2011}, where it is derived for the signal-plus-background experiment. For $\delta = 0$, the distribution reduces to the chi-bar-square, and for $\delta \to \infty$, it reduces to the chi-square distribution.

An example from $X \sim \mathcal{N}(\mu, 1)$ with $\mu\in M_1$ is shown in Figure~\ref{fig:cc_constrained_example}. The left panel shows three different confidence curves from the observation $x=-1$: one based on the naïve MLE $\tilde{\mu}(X)$ and the $\chi^2_1$ distribution and two based on the proper MLE $\hat{\mu}(X)$, hereafter termed \emph{boundary-aware} confidence curves. Of the latter two, one is incorrectly constructed using the $\chi^2_1$ distribution, the other uses the correct distribution derived by \cite{WuNeale2012}. The confidence curve from the naïve approach is the same as the one referred to as \emph{classical} in \cite{Mandelkern2002}. \citet[Table 1]{Mandelkern2002} shows that the classical 68\% confidence interval from this setting is empty, which is also observed in the left panel of Figure \ref{fig:cc_constrained_example}. However, the correctly specified boundary-aware confidence curve gives a 68\% confidence interval of $[0, 0.267]$. The true parameter in this setting was set to $\mu_0 = 0.5$, and the right panel of Figure~\ref{fig:cc_constrained_example} shows the empirical coverage of the three confidence curves across all levels $1-\alpha$, based on $10^5$ MC simulated observations $x$. Both the naïve and the boundary-aware confidence curve from Eq.~\eqref{eq:CDF-univariate} follow the nominal $U(0,1)$ line and are therefore correctly calibrated. In contrast, the boundary-aware confidence curve constructed using the incorrect $\chi^2_1$ distribution overcovers in this close-to-boundary example.

\subsection{Two-dimensional parameter case}\label{sec:bivar-formulation}
The bivariate case has no counterpart to the univariate result of \citet{WuNeale2012}. Two bounded parameters are, however, often of interest simultaneously; in the context of high-energy physics see, for instance, \citet{athron_global_2017, athron_status_2017, ananyev_collider_2023}. In this case, the natural summaries are confidence regions, or more generally, bivariate confidence curves. 

\subsubsection{Framework}
Consider $X = (X_1, X_2) \sim \mathcal{N}_2(\mu, \Sigma)$, where $\mu \in M_2$ with unknown true values $\mu_0 = (\mu_{0,1}, \mu_{0,2})$, and known covariance 
$$\Sigma = \begin{pmatrix} \sigma_1^2 & \rho\sigma_1\sigma_2 \\ \rho\sigma_1\sigma_2 & \sigma_2^2 \end{pmatrix}$$
where $\rho \in (-1, 1)$ is the correlation coefficient. The parameter space is $M_2 = \Omega_1 \times \Omega_2$, where each component $\mu_i$ is either bounded below at zero ($\Omega_i=[0, \infty)$) or has no boundary ($\Omega_i = \mathbb{R}$). The likelihood ratio statistic for testing a hypothesised value $\mu \in M_2$ is
\begin{equation*}
\begin{split}
    q(\mu; X) &= (X - \mu)^\top \Sigma^{-1} (X - \mu) \\
    &-\left\{X - \hat{\mu}(X)\right\}^\top \Sigma^{-1} \left\{X - \hat{\mu}(X)\right\},
\end{split}
\end{equation*}
where the MLE $\hat{\mu}(X)$ is found by solving
\begin{equation}
    \hat{\mu}(X)=\arg\min_{\mu\in M_2}(X-\mu)^{\top}\Sigma^{-1}(X-\mu),
    \label{eq:MLE-bivar}
\end{equation}
in which the level sets of the objective function, $(X-\mu)^{\top}\Sigma^{-1}(X-\mu)$, are ellipses centred at the naïve MLE $\tilde{\mu}(X) = X$, as shown in Figures~\ref{fig:bivar-1bounded} and~\ref{fig:2d-2params}. If $M_2=\mathbb{R}^2$, then automatically $\hat{\mu}(X) = X$ and the likelihood ratio statistic reduces to a single quadratic form, which is distributed as a $\chi^2_2$ under the true parameter. When $M_2$ is a proper subset of $\mathbb{R}^2$, however, the naïve MLE $\tilde{\mu}(X)$ has positive probability of falling outside the parameter space. In that case, $\hat{\mu}(X)$ is the projection of $\tilde{\mu}(X)$ onto $M_2$ in the metric induced by the precision matrix $\Sigma^{-1}$, or geometrically, the point where the smallest level set ellipse is tangent to $M_2$. This projection depends on the entries of the precision matrix, which we write as
\begin{equation*}
    \Sigma^{-1} = \frac{1}{\sigma_1^2\sigma_2^2(1-\rho^2)}\begin{pmatrix} \sigma_2^2 & -\rho\sigma_1\sigma_2 \\ -\rho\sigma_1\sigma_2 & \sigma_1^2 \end{pmatrix}=\begin{pmatrix} a & b \\ b & c \end{pmatrix}.
\end{equation*}
For more details, see \citet[Section~3.12]{silvapulle_constrained_2005}.

In order to construct confidence regions and bivariate confidence curves for $\mu$ with correct coverage, we need the distribution of the statistic under each null candidate $\mu \in M_2$. This requires the joint distribution of the statistic within each projection region and the probability of that region. The total CDF of the likelihood ratio statistic can be written as
\begin{equation}
    F_{q}(t;\mu)
    = \sum_j \operatorname{Pr}_\mu\left\{q_j(\mu; X) \leq t,\; \tilde{\mu}(X) \in C_j\right\},
    \label{eq:branch-decomp-biv}
\end{equation}
where the sets $C_j$ partition the naïve parameter space ($\mathbb{R}^2$) according to which projection region $\tilde{\mu}(X)$ falls into, where on each $C_j$ the likelihood ratio statistic takes a different form, $q_j(\mu; X)$. We refer to these $C_j$'s as \emph{branches} in the remaining text.

\subsubsection{One bounded dimension}\label{sec:2param_1bounded}

\begin{figure}
    \centering
    \includegraphics[width=0.52\linewidth]{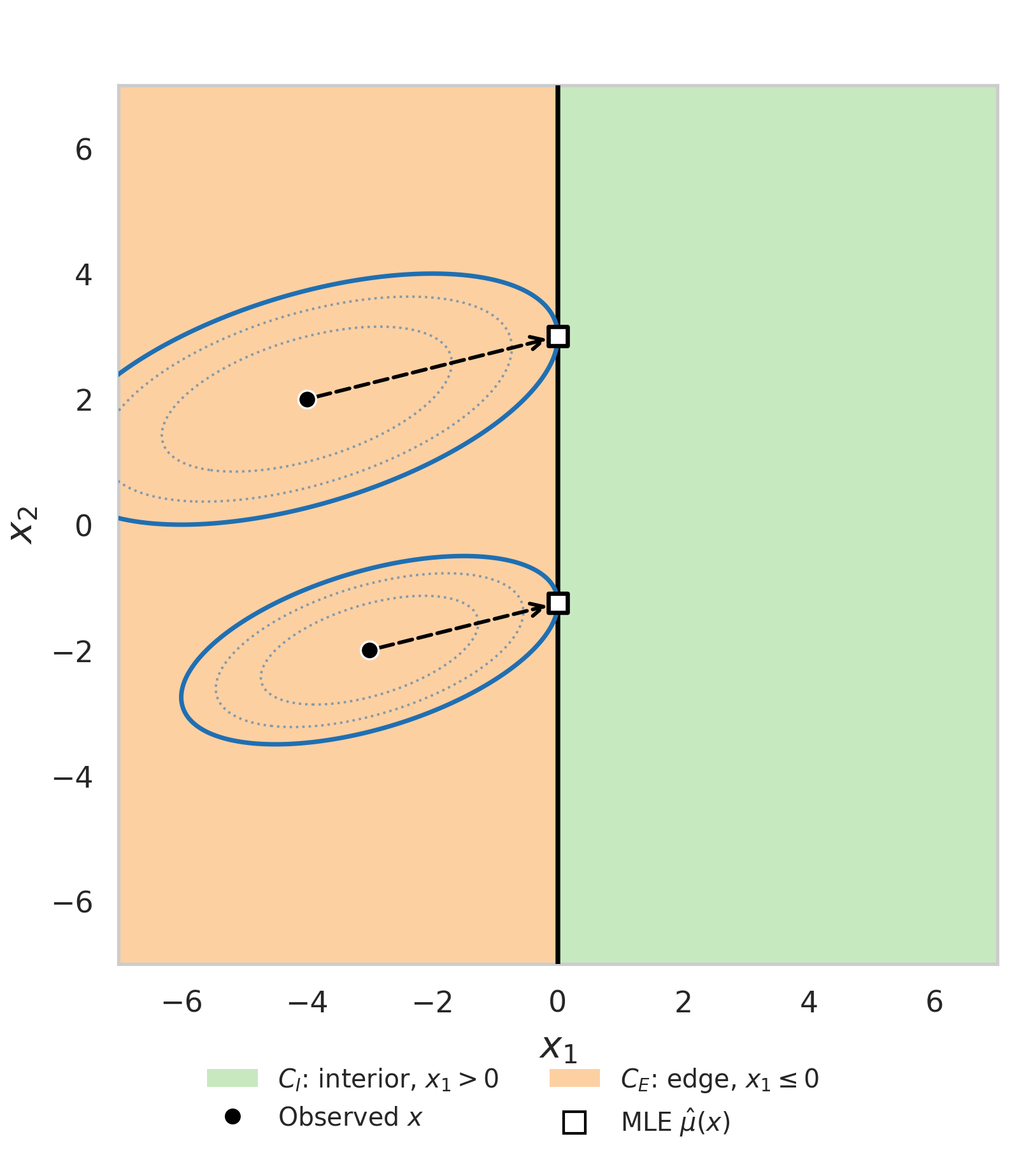}
    \caption{The two branches $C_I$ and $C_E$ for the bivariate normal with one bounded component: $\mu \in [0, \infty) \times \mathbb{R}$, variances $\sigma_1^2=4$, $\sigma_2^2=1$, and correlation $\rho = 0.5$. Dashed arrows show the transformations from two observations $x$ (black dots) to their MLEs $\hat{\mu}(x)$ on the boundary face, as determined by the $\Sigma^{-1}$-metric.}
    \label{fig:bivar-1bounded}
\end{figure}
Let $M_2=[0,\infty)\times\mathbb{R}$. The MLE, found by solving Eq.~\eqref{eq:MLE-bivar}, is
\begin{equation}
    \hat{\mu}(X)
    =
    \begin{cases}
         X &  X_1>0, \\[1ex]
        \left(0,\ X_2+\frac{b}{c} X_1\right)^\top
        & X_1\leq 0.
    \end{cases}
    \label{eq:MLE-one-bounded}
\end{equation}
Figure~\ref{fig:bivar-1bounded} shows the projections of two observations $x$ of $X$ which make $\tilde{\mu}(X)$ fall outside $M_2$. Each is projected onto the corresponding MLE $\hat{\mu}(x)$ according to Eq.~\eqref{eq:MLE-one-bounded}, as visualised by the dashed arrows. The CDF of the likelihood ratio statistic decomposes into two branches,
\begin{equation*}
\begin{split}
        F_q(t; \mu) &= \operatorname{Pr}_\mu\{q(\mu; X) \leq t, X_1 > 0\} \\
        &+ \operatorname{Pr}_\mu\{q(\mu; X) \leq t, X_1 \leq 0\}.
\end{split}
\end{equation*}

Let us transform $X$ to the coordinates $(V,U)$, which are independent standard normal random variables defined by
\begin{equation}
\begin{split}
&V = \frac{X_1-\mu_1}{\sigma_1}, \\
&U = \frac{bX_1 + cX_2 - (b\mu_1 + c\mu_2)}{\sqrt{c}}.
\end{split}
\label{eq:VandU}
\end{equation}
Additionally, let $\delta=\mu_1/\sigma_1$ denote the standardised distance to the boundary. Since $X_1=\mu_1+\sigma_1V$, the events $X_1>0$ and $X_1\leq 0$ are equivalent to $V>-\delta$ and $V\leq-\delta$, respectively. The likelihood ratio statistic can be written as $q(\mu;X)=U^2+B_\delta(V)$, where $B_\delta(V)$ is the contribution of the bounded coordinate,
\begin{equation*}
    B_\delta(V)=
    \begin{cases}
    V^2, & V>-\delta, \\[4pt]
    -\delta^2-2\delta V, & V\leq-\delta.
    \end{cases}
\end{equation*}
The term $B_\delta(V)$ is the same likelihood ratio statistic as for the one-dimensional bounded mean in Section~\ref{sec:1d-WuNeale}, which alone gives the CDF in Eq.~\eqref{eq:CDF-univariate}. The complete likelihood ratio statistic, however, also contains the independent contribution $U^2$ from the unbounded coordinate. Thus, its distribution is not simply the one-dimensional bounded distribution, but rather the convolution of that distribution with a $\chi^2_1$ variable.

For a fixed value $V=v$, the likelihood ratio statistic satisfies $q(\mu;X)\leq t$ if and only if $U^2\leq t-B_\delta(v)$. Since $U^2\sim\chi^2_1$ independently of $V$, the cumulative distribution function of the likelihood ratio statistic is
\begin{equation*}
F_{q}(t; \mu)
=
\int_{-\infty}^{\infty}
F_{\chi^2_1}\{t-B_\delta(v)\}\phi(v) dv,
\end{equation*}
where $\phi$ is the standard normal density. The convolution integral can be split according to the two branches of $B_\delta$. This gives
\begin{equation}
\begin{split}
    F_{q}(t; \mu)
    &=
    \int_{\max(-\delta, -\sqrt{t})}^{\sqrt{t}}
    F_{\chi^2_1}\left\{t-v^2\right\}\phi(v)\,dv \\
    &+
    \int_{-(t+\delta^2)/(2\delta)}^{-\delta}
    F_{\chi^2_1}\left\{t+\delta^2+2\delta v\right\}\phi(v)\,dv, \label{eq:CDF_2dim_1bounded}
\end{split}
\end{equation}
where the integration bounds are determined by the branch conditions: $V>-\delta$ for the first branch and $V\leq -\delta$ for the second, together with $0 \le U^2 \le t - B_\delta(v)$, which requires $t - B_\delta(v) \ge 0$. In general, Eq.~\eqref{eq:CDF_2dim_1bounded} does not have a direct expression in terms of standard statistical functions, such as the standard normal distribution functions in Eq.~\eqref{eq:CDF-univariate}.

The two limits, as expected, recover familiar laws. As $\delta\to\infty$ the null lies far from the boundary, so $\operatorname{Pr}_\mu(V\le-\delta)\to0$: the second integral vanishes and $F_q(t;\mu)\to F_{\chi^2_2}(t)$, recovering Wilks' theorem. At $\delta=0$ the null lies on the boundary, where the bounded coordinate contributes $V^2$ when $V>0$ and nothing when $V\le0$; the two branches then reduce to the $0.5\chi^2_1 + 0.5\chi^2_2$ chi-bar-square distribution of \citet[Case 6]{SelfLiang1987}.

\subsubsection{Both dimensions bounded}\label{sec:2param_2bounded}
\begin{figure}[t]
    \centering
    \includegraphics[width=0.52\linewidth]{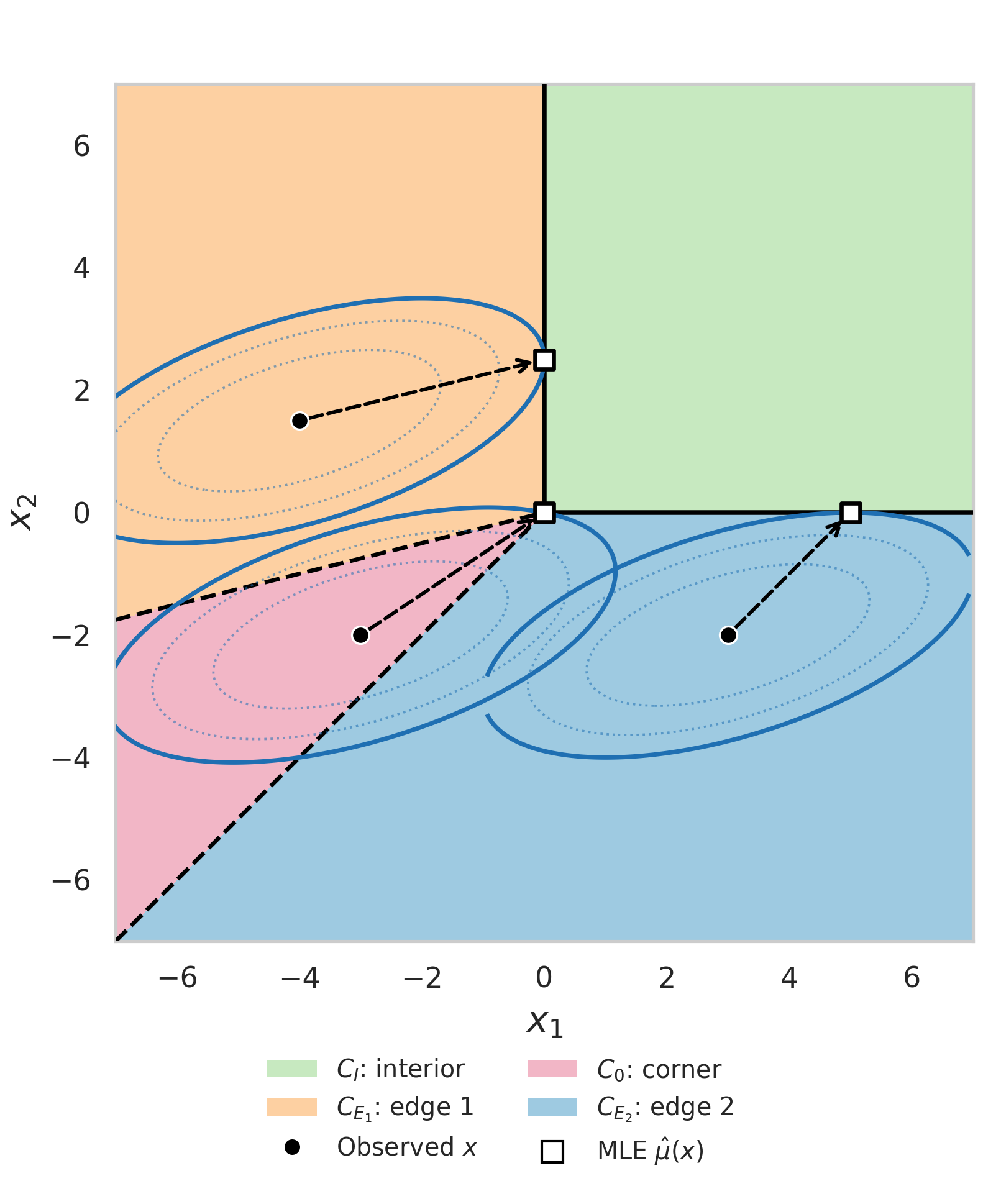}
    \caption{The four branches $C_I$, $C_{E_1}$ and $C_{E_2}$, and $C_0$ for the bivariate normal with both components bounded: $\mu \in [0,\infty)^2$, variances $\sigma_1^2 = 4$, $\sigma_2^2 = 1$, and correlation $\rho = 0.5$. The dashed arrows show the transformations from three observations $x$ (black dots) to their respective MLEs $\hat{\mu}(x)$: to one of the two boundary faces from $C_{E_1}$ and $C_{E_2}$, and to the origin from $C_0,$ as determined by the $\Sigma^{-1}$-metric.}
    \label{fig:2d-2params}
\end{figure}
Let $M_2 = [0,\infty)\times[0,\infty)$, i.e., the non-negative quadrant. Solving Eq.~\eqref{eq:MLE-bivar} leaves four possibilities for where on $M_2$ the MLE is projected to from $\tilde{\mu}(X)$, depending on where $\tilde{\mu}(X)$ lies. The four possibilities are:  an interior point; projection onto each of the two boundary faces; and projection onto the corner at the origin. We describe these cases via four branches $C_I$, $C_{E_1}$, $C_{E_2}$, and $C_0$, visualised in Figure~\ref{fig:2d-2params}. The corresponding MLE and likelihood ratio forms are summarised in Table~\ref{tab:branches-twobounded}.

\begin{table*}[t]
\centering
\caption{The four branches $C_j$, formula for their MLE $\hat{\mu}_j(X)$ and likelihood ratio $q_j(\mu; X)$ for the bivariate normal model where $M_2 = [0,\infty)\times[0,\infty)$.\\
}
\label{tab:branches-twobounded}
\begin{tabular}{lll}
\hline
\rule{0pt}{2.5ex} Branch $C_j$ & $\hat\mu_j(X)$ & Likelihood ratio form $q_j(\mu;X)$ \\[0.3em]
\hline
\rule{0pt}{2.5ex}
$C_I$ &
$X$ &
$(X-\mu)^\top \Sigma^{-1} (X-\mu)$
\\[0.4em]
$C_{E1}$ &
$\displaystyle
\left(
0,
X_2 + X_1 b/c\right)^\top$ &
$\displaystyle
(X-\mu)^\top \Sigma^{-1} (X-\mu)
- X_1^2(ac-b^2)/c $
\\[0.8em]
$C_{E2}$ &
$\displaystyle
\left(
X_1 + X_2 b/a,
0
\right)^\top$ &
$\displaystyle
(X-\mu)^\top \Sigma^{-1} (X-\mu)
- X_2^2(ac-b^2)/a$
\\[0.8em]
$C_0$ &
$(0,0)^\top$ &
$\displaystyle
(X-\mu)^\top \Sigma^{-1} (X-\mu)
- X^\top \Sigma^{-1} X$
\\[0.4em]
\hline
\end{tabular}
\end{table*}

The interior branch $C_I$ collects those observations for which the naïve MLE already lies in $M_2$. The two edge branches, $C_{E_1}$ and $C_{E_2}$, correspond to observations that project the naïve MLE onto the relative interiors of the boundary faces $\{\mu_1 = 0, \mu_2 > 0\}$ and $\{\mu_1 > 0, \mu_2 = 0\}$, respectively. Finally, $C_0$ is the corner branch for which the naïve MLE is projected to the origin. The CDF of the likelihood ratio statistic decomposes accordingly as
\begin{equation}
    F_q(t;\mu)
    =
    I_I(t;\mu)
    +
    I_{E_1}(t;\mu)
    +
    I_{E_2}(t;\mu)
    +
    I_0(t;\mu),
    \label{eq:2bounded-total-cdf}
\end{equation}
where $I_j(t;\mu)$ is the contribution from branch $C_j$.

\paragraph{Interior branch} Here, the likelihood ratio statistic reduces to the usual squared Mahalanobis distance from the observation to the hypothesised $\mu$. The contribution $I_I(t;\mu)$ is therefore the probability that the naïve MLE $\tilde{\mu}(X)$ falls in the intersection of the parameter space $M_2$ and the ellipse $q_I(\mu;X) \leq t$.

Let us transform $X$ into $Y = L^{-1}(X-\mu)$, where $L$ is the lower-triangular Cholesky factor of $\Sigma = LL^\top$, so that $Y \sim \mathcal{N}_2(0, I_2)$ with $I_2$ the two-dimensional identity matrix. In polar coordinates, $Y = R \omega(\phi)$, where $R = \|Y\|$ and $\omega(\phi) = (\cos\phi, \sin\phi)$ for $\phi \in [0, 2\pi)$. The likelihood ratio statistic is then $q(\mu;X) = R^2$, so that $q(\mu; X) \leq t$ is equivalent to $R \leq \sqrt{t}$. Moreover, the interior constraint becomes $\mu + LR\omega(\phi) \geq 0$. Let $A_I(r)$ be the total angle spanned by the $\phi$ that satisfy this interior constraint for a fixed $r$. Then, 
\begin{equation}
    I_I(t; \mu) = \int_0^{\sqrt{t}} f_Y(r) A_I(r) r dr,
    \label{eq:interior_integral}
\end{equation} 
where $f_Y(r) = e^{-r^2/2}/(2\pi)$ is the density of $Y$ written in terms of $r$. The function $A_I(r)$ is not trivial to compute. If the null is placed at the corner of $M_2$, i.e.\ $\mu = (0,0)$, $A_I(r)$ is the constant $\pi/2 + \arcsin\rho$. The radial integral then factorises as $I_I(t; 0) = w_I\, F_{\chi^2_2}(t)$ with interior weight $w_I = 0.25 + \arcsin(\rho)/(2\pi)$, recovering the weight of the first term in \citet[Case 7]{SelfLiang1987}, which corresponds to the bivariate orthant probability $\operatorname{Pr}_{\mu=0}(X_1\geq 0, X_2\geq 0)$. For a general value of $\mu$, however, this angle must be computed by
\begin{equation*}
    A_I(r) = \int_0^{2\pi} \mathbf{1}\{\mu + Lr \omega(\phi)\geq 0\} d\phi,
\end{equation*}
where $\mathbf{1}\{\cdot\}$ is the indicator function.

\paragraph{Edge branches} Here, the MLE is obtained by projecting the naïve MLE $\tilde{\mu}(X)$ onto the corresponding boundary face. First, let us consider the first edge branch $C_{E_1}$, noting that the result transfers to the other edge branch by switching the coordinates.

In order to derive the analytic distribution, we transform $X$ into the independent standard normal coordinates $(U, V)$ as in Eq.~\eqref{eq:VandU}. In contrast to the one bounded parameter case, the edge branch forces constraints on both $U$ and $V$. Specifically, $C_{E_1} = \{ (U,V) : U > -\gamma,\ V \leq -\delta \}$, where 
$$\gamma = \frac{b\mu_1 + c\mu_2}{\sqrt{c}}, \quad \delta = \frac{\mu_1}{\sigma_1}.$$

On this branch the likelihood ratio statistic can be written as $$q_{E_1}(\mu; X) = U^2 - \delta^2 - 2\delta V.$$ Fixing $U = u$, the event $\{q(\mu; X) \le t\}$ defines a lower bound on $V$ which, together with the edge constraint $V \le -\delta$, restricts $V$ to a finite interval. Since $V$ is standard normal, the inner probability reduces to a difference of $\Phi$–terms. Integrating over $u$, subject to the edge bound $u > -\gamma$ and to the interval being non-empty, yields the one-dimensional integral
\begin{equation}
\begin{split}
    I_{E_1}(t; \mu) &= \int_{\max(-\gamma,\, -\sqrt{t-\delta^2})}^{\sqrt{t-\delta^2}} \phi(u) \\
    &\times \left\{\Phi(-\delta) - \Phi\!\left(\frac{u^2 - \delta^2 - t}{2\delta}\right)\right\} du,
    \label{eq:edge1_integral}
\end{split}
\end{equation}
where we have integrated over $V$ analytically since it enters $q(\mu; X)$ linearly; the bound on $U$ is then absorbed into the outer limits. 

The second edge branch $C_{E_2}$, the face $\{\mu_1 > 0, \mu_2 = 0\}$, follows by interchanging the roles of the two coordinates. Define
\begin{equation}
\begin{split}
    V' &= \frac{X_2 - \mu_2}{\sigma_2}, \quad U' = \frac{aX_1 + bX_2 - (a\mu_1 + b\mu_2)}{\sqrt{a}},\\ \gamma' &= \frac{a\mu_1 + b\mu_2}{\sqrt{a}}, \quad \delta' = \frac{\mu_2}{\sigma_2}. \label{eq:VandUprime}
\end{split}
\end{equation}
The contribution $I_{E_2}(t;\mu)$ is then given by Eq.~\eqref{eq:edge1_integral} with ($\delta$, $\gamma$) replaced by ($\delta'$, $\gamma'$).

The contribution of both edge branches to the total CDF in Eq.~\eqref{eq:2bounded-total-cdf} vanishes in the deep interior ($\delta, \delta' \to \infty$), as the probability of $X$ landing inside any of the two edge branches is zero. On the opposite extreme, placing the null at the corner, $\mu = (0,0)$, gives $\delta = \delta' = \gamma = \gamma' = 0$. The likelihood ratio on the first edge then reduces to $q(\mu;X) = U^2$ with branch conditions $U > 0$ and $V \leq 0$, so the two variables separate and
$$
I_{E_1}(t;0) = I_{E_2}(t;0) = \frac{1}{2}\int_{0}^{\sqrt{t}}\phi(u)\,du = \frac{1}{4}F_{\chi^2_1}(t),
$$
independently of $\rho$, in accordance with \citet[Case 7]{SelfLiang1987}.
\paragraph{Corner branch} Here, the MLE is obtained by projecting the naïve MLE $\tilde{\mu}(X)$ to the origin. Let 
$$W = \Sigma^{-1}(X-\mu), \quad \lambda = \Sigma^{-1}\mu, \quad \kappa = \mu^\top \Sigma^{-1} \mu,$$
such that $W \sim \mathcal{N}_2(0, \Sigma^{-1})$. The corner event becomes $\{X \in C_0\} = \{W_1 \leq -\lambda_1, W_2 \leq -\lambda_2\}$, and the likelihood ratio of the corner branch can be written as 
$$q_0(\mu;X) = -2\mu^\top W - \kappa.$$
Hence, $q(\mu;X) \leq t$ is equivalent to $-\mu^\top W \leq (t + \kappa)/2$, and the corner contribution can be written as 
$$I_0(t;\mu) = \operatorname{Pr}_\mu\left( W_1\leq -\lambda_1,\, W_2\leq -\lambda_2,\, -\mu^\top W\leq \frac{t+\kappa}{2} \right).$$
This is the joint probability of three linear functionals of the same two-dimensional Gaussian vector $W$. Defining $Z=(W_1,W_2,-\mu^\top W)^\top=BW$, with
$$B= \begin{pmatrix} 1 & 0\\ 0 & 1\\ -\mu_1 & -\mu_2 \end{pmatrix},$$
we have $Z\sim\mathcal{N}_3(0,B\Sigma^{-1}B^\top)$. The covariance matrix $B\Sigma^{-1}B^\top$ has rank two, and is therefore singular, because $B$ maps the two-dimensional vector $W$ into a two-dimensional subspace of $\mathbb{R}^3$. Hence the corner contribution can be expressed as the singular trivariate normal CDF 
\begin{equation} 
I_0(t;\mu) = \Phi_3\left\{ \begin{pmatrix} -\lambda_1 \\ -\lambda_2 \\ (t+\kappa)/2 \end{pmatrix}; 0,\, B\Sigma^{-1}B^\top \right\}. 
\label{eq:corner_CDF} 
\end{equation}

On the corner, $\mu = (0,0)$, both $\lambda$ and $\kappa$ vanish and the third constraint reduces to $0 \leq t/2$, which is always satisfied. Eq.~\eqref{eq:corner_CDF} then collapses to the orthant probability $\operatorname{Pr}(W_1 \leq 0, W_2 \leq 0)$, contributing the point mass $w_0 = 0.25 -\arcsin\rho/(2\pi)$ at $t = 0$ and recovering the $\chi^2_0$ term of the chi-bar-square limit of \citet[Case 7]{SelfLiang1987}. As $\mu$ moves into the interior of $M_2$, the thresholds $-\lambda_1, -\lambda_2 \to -\infty$ and so $I_0(t;\mu) \to 0$, meaning the corner branch becomes negligible.

\section{Chi-bar-square approximation }\label{sec:calibration}
The branch decompositions developed in Sections~\ref{sec:2param_1bounded} and~\ref{sec:2param_2bounded} express the full null CDF of the likelihood ratio statistic as a sum of the individual CDFs from each branch. Apart from the corner branch of Eq.~\eqref{eq:corner_CDF}, which can be evaluated with standard routines, each branch requires numerical quadrature of a non-standard integral. These integrals are inexpensive to evaluate but might be awkward to use. We therefore develop a surrogate in terms of standard statistical functions, which approximates the distribution of the statistic within each branch by a simpler one. 

\subsection{Moment-matching}\label{sec:moment-matching}
Partitioning $\mathbb{R}^2$ into the branches $C_j$, the null distribution can be written as
\begin{equation*}
  F_q(t) = \sum_j w_j \operatorname{Pr}_\mu\!\big(q_j(\mu;X) \le t \mid \tilde{\mu}(X) \in C_j\big).
\end{equation*}
The weights $w_j = \operatorname{Pr}_\mu(C_j)$ are Gaussian orthant probabilities and are directly available in terms of standard statistical functions for every branch considered here. Therefore, only the conditional laws need to be approximated. On each branch we match the conditional law to a shifted and scaled chi-square,
\begin{equation*}
  \operatorname{Pr}_\mu\!\left(q_j(\mu;X) \leq t \mid \tilde{\mu}(X) \in C_j\right)
  \approx F_{\chi^2_{\nu_j}}\!\left(\frac{t - s_j}{c_j}\right),
\end{equation*}
where the shift $s_j$ is the deterministic lower bound of $q_j(\mu;X)$ on the branch $j$ and the pair $(c_j,\nu_j)$ is fixed by the first two conditional moments $m_j$ and $v_j$, as $c_j = v_j/\{2(m_j-s_j)\}$ and $\nu_j = 2(m_j-s_j)^2/v_j$. In the interior branch, the statistic is a genuine quadratic form and $s_j=0$; the shift is needed only on the edge branches, where projection onto the boundary faces displaces the statistic away from the origin. Collecting the branches gives
\begin{equation}
  F_q(t) \approx \sum_j w_j \mathbf{1}\{t \ge s_j\} F_{\chi^2_{\nu_j}}\Big(\tfrac{t-s_j}{c_j}\Big),
  \label{eq:chibar-approx}
\end{equation}
a finite mixture of shifted chi-squares that reduces to the chi-bar-square law on the boundary and to a single chi-square in the deep interior.

\begin{table}[t]
\centering
\caption{Branchwise moment-matching components for the bivariate normal with two bounded mean parameters: $M_2 = [0,\infty)\times[0,\infty)$.}
\label{tab:branches-moments}
\vspace{0.5em}
\makebox[\textwidth][c]{%
\begin{tabular}{llllll}
\hline
\rule{0pt}{2.5ex}Branch $C_j$ & $(q_j(\mu;X) \mid X\in C_j) \approx$ & Mean $m_j$ & Variance $v_j$ & Shift $s_j$ & Weight $w_j$ \\[0.3em]
\hline
\rule{0pt}{2.5ex}$C_I$ &
$A(\delta) + A(\delta')$ &
$m_A(\delta) + m_A(\delta')$ &
$v_A(\delta) + v_A(\delta')$ &
$0$ &
$\Phi_2(\delta, \delta'; \rho)$
\\[0.4em]
$C_{E_1}$ &
$A(\gamma) + B(\delta)$ &
$m_A(\gamma) + m_B(\delta)$ &
$v_A(\gamma) + v_B(\delta)$ &
$\delta^2$ &
$\Phi(-\delta)\Phi(\gamma)$
\\[0.4em]
$C_{E_2}$ &
$A(\gamma') + B(\delta')$ &
$m_A(\gamma') + m_B(\delta')$ &
$v_A(\gamma') + v_B(\delta')$ &
$\delta'^2$ &
$\Phi(-\delta')\Phi(\gamma')$
\\[0.4em]
$C_0$ &
exact, Eq.~\eqref{eq:corner_CDF} &
--- &
--- &
--- &
---
\\[0.4em]
\hline
\end{tabular}%
}
\end{table}

\subsection{One bounded dimension}
For the two-dimensional mean with one of the dimensions bounded, $\mathbb{R}^2$ splits into an interior branch $\{V>-\delta\}$, in which the likelihood ratio is $q_I(\mu; X)=U^2+V^2$, and an edge branch $\{V\le-\delta\}$, in which $q_E(\mu; X)=U^2-\delta^2-2\delta V$. The weights are $w_I=\Phi(\delta)$ and $w_E=\Phi(-\delta)$, the interior statistic is unshifted, and the edge statistic is bounded below by $\delta^2$. The conditional moments follow from truncated standard normal moments. The unbounded coordinate $U$ enters both branches only through its square and is independent of the truncation, and contributes $1$ to the mean and $2$ to the variance. On the interior branch $V$ is truncated below at $-\delta$. Writing $\lambda_+ = \phi(\delta)/\Phi(\delta)$, integration by parts gives $E[V^2 \mid V>-\delta] = 1-\delta\lambda_+$ and $E[V^4 \mid V>-\delta] = 3-(\delta^3+3\delta)\lambda_+$, so that
\begin{equation*}
\begin{split}
  m_I &= 2 - \delta\lambda_+, \\
  v_I &= 5 - (\delta^3+3\delta)\lambda_+ - \{1-\delta\lambda_+\}^2.
\end{split}
\end{equation*}
On the edge branch, $V$ is truncated above at $-\delta$. Writing $\lambda_- = \phi(\delta)/\Phi(-\delta)$, we have $E[V \mid V\le-\delta] = -\lambda_-$ and $\mathrm{Var}(V \mid V\le-\delta) = 1+\delta\lambda_--\lambda_-^2$, and since $q_E(\mu; X)$ is linear in $V$,
\begin{equation*}
\begin{split}
  m_E &= 1 - \delta^2 + 2\delta\lambda_-, \\
  v_E &= 2 + 4\delta^2\{1+\delta\lambda_--\lambda_-^2\}.
\end{split}
\end{equation*}
With shifts $s_I=0$ and $s_E=\delta^2$, the scale and degrees of freedom on each branch follow from Section~\ref{sec:moment-matching} as $c_j=v_j/\{2(m_j-s_j)\}$ and $\nu_j=2(m_j-s_j)^2/v_j$. Substituting these expressions into Eq.~\eqref{eq:chibar-approx} gives
\begin{equation*}
  F_q(t)
  \approx
  w_I F_{\chi^2_{\nu_I}}\left(\frac{t}{c_I}\right)
  +
  w_E \mathbf{1}{\{t \ge \delta^2\}}
        F_{\chi^2_{\nu_E}}\left(\frac{t - \delta^2}{c_E}\right).
\end{equation*}
The approximation is exact in both limits: as $\delta\to\infty$ the edge weight vanishes and the interior moments return $(m_I,v_I)=(2,4)$, hence $(c_I,\nu_I)=(1,2)$, recovering the $\chi^2_2$ law, while at $\delta=0$ the moments give $(c_I,\nu_I)=(1,2)$ and $(c_E,\nu_E)=(1,1)$ with zero shift, with weights $w_I = w_E = \Phi(0) = 0.5$, so the two branches reduce to the $0.5\chi^2_1+0.5\chi^2_2$ mixture.

\subsection{Both dimensions bounded}\label{sec:mm2}
With two parameters bounded, there are four branches: the interior, two edges and a corner. The corner's contribution is a simple form (the singular trivariate normal probability in Eq.~\eqref{eq:corner_CDF}) so the approximation is applied only to the interior and the two edges. On these latter branches the conditional statistic decomposes, coordinate by coordinate, into two univariate building blocks. The first block is a square coordinate that satisfies its constraint,
\begin{equation}
  A(a) \;=\; Z^2 \,\big|\, Z>-a, \qquad Z\sim \mathcal{N}(0,1),
  \label{eq:blockA}
\end{equation}
with moments,
\begin{equation*}
\begin{split}
  m_A(a) &= 1 - a\lambda_+(a), \\
  v_A(a) &= 3 - (a^3+3a)\lambda_+(a) - \{1-a\lambda_+(a)\}^2,
\end{split}
\end{equation*}
where $\lambda_+(a)=\phi(a)/\Phi(a)$ is the inverse Mills ratio. The second block is the shifted linear contribution of a coordinate projected onto its constraint,
\begin{equation}
  B(\delta) \;=\; -\delta^2 - 2\delta Z \,\big|\, Z\le-\delta, \qquad Z\sim \mathcal{N}(0,1),
  \label{eq:blockB}
\end{equation}
with support $[\delta^2,\infty)$ and moments
\begin{equation*}
\begin{split}
  m_B(\delta) &= -\delta^2 + 2\delta\lambda_-(\delta), \\
  v_B(\delta) &= 4\delta^2\{1+\delta\lambda_-(\delta)-\lambda_-^2(\delta)\},
  \end{split}
\end{equation*}
where $\lambda_-(\delta)=\phi(\delta)/\Phi(-\delta)$.

Table~\ref{tab:branches-moments} shows how the branches relate to these blocks. On each edge branch the decomposition is exact. In the first edge branch, we transform $(X_1, X_2)$ to the independent coordinates $(U, V)$ of Eq.~\eqref{eq:VandU}. The $U$ coordinate contributes $A(\gamma)$ and $V$ contributes $B(\delta)$. The conditional distributions are independent, so the conditional moments are $m_{E_1}=m_A(\gamma)+m_B(\delta)$ and $v_{E_1}=v_A(\gamma)+v_B(\delta)$, with shift $s_{E_1}=\delta^2$. The second edge follows by substituting $(\delta,\gamma)$ with $(\delta',\gamma')$. On the interior branch, the statistic is the squared Mahalanobis distance from $X$ to $\mu$, and the branch is the event $\{V > -\delta,\ V' > -\delta'\}$, with $V$ and $V'$ as in Eqs.~\eqref{eq:VandU} and \eqref{eq:VandUprime}. We approximate its conditional law as if $V$ and $V'$ were independent, yielding $A(\delta) + A(\delta')$. This is exact for $\rho = 0$ but for $\rho \neq 0$ it ignores the correlation in the conditional shape, while the weight $w_I = \Phi_2(\delta, \delta'; \rho)$ remains exact. Exact interior moments are available. However, these entail long expressions for the conditional means and variances, making them, in our opinion, less favourable than the integral form in Eq.~\eqref{eq:interior_integral}. We evaluate the performance of the simplified interior moments in the following section. 

Given the entries of Table~\ref{tab:branches-moments}, the scale and degrees of freedom on each branch follow from the moment-matching equations of Section~\ref{sec:moment-matching}: $c_j=v_j/\{2(m_j-s_j)\}$ and $\nu_j=2(m_j-s_j)^2/v_j$. Collecting the four contributions, we obtain
\begin{equation}
\begin{split}
  F_q(t) &\approx w_I F_{\chi^2_{\nu_I}}\left(t/c_I\right) \\
  &+ w_{E_1} \mathbf{1}\left\{t\ge\delta^2\right\} F_{\chi^2_{\nu_{E_1}}}\left\{(t-\delta^2)/c_{E_1}\right\} \\
  &+ w_{E_2} \mathbf{1}\left\{t\ge\delta'^2\right\} F_{\chi^2_{\nu_{E_2}}}\left\{(t-\delta'^2)/c_{E_2}\right\} \\
  &+ I_0(t;\mu),
\end{split}
\label{eq:approx-twobounded}
\end{equation}
with $I_0$ being the exact corner term. All ingredients are elementary functions of the standardised boundary distances $(\delta,\delta',\gamma,\gamma')$ together with one bivariate orthant probability, so Eq.~\eqref{eq:approx-twobounded} requires no numerical integration beyond the evaluation of the chi-square and normal CDFs. 

In the two extremes, $\mu=0$ and $\mu\to\infty$, the moment-matched approximation reduces to known forms. When the null is on the corner of the parameter space, all four boundary distances vanish, $\delta=\delta'=\gamma=\gamma'=0$, and the inverse Mills ratios take the value $\lambda_+(0)=\lambda_-(0)=\sqrt{2/\pi}$. The blocks of Eqs.~\eqref{eq:blockA} and \eqref{eq:blockB} then carry the moments $m_A(0)=1$, $v_A(0)=2$ and $m_B(0)=0$, $v_B(0)=0$, respectively. Following Table~\ref{tab:branches-moments}, Eq.~\eqref{eq:approx-twobounded} then coincides with the chi-bar-square $w_0\chi^2_0+0.5\chi^2_1+w_I\chi^2_2$ of \citet[Case 7]{SelfLiang1987}. In the opposite limit, $\delta,\delta', \gamma, \gamma'\to\infty$, the edge and corner weights are zero, while the interior carries all the weight, i.e.\ $w_I=1$, and the moments $m_A=1$, $v_A=2$, such that $(c_I,\nu_I)\to(1,2)$. The mixture then collapses to a single $\chi^2_2$-distribution, recovering Wilks' theorem.

\begin{figure}[htbp]
    \centering
    \makebox[\textwidth][c]{%
        \includegraphics[width=1.5\textwidth]{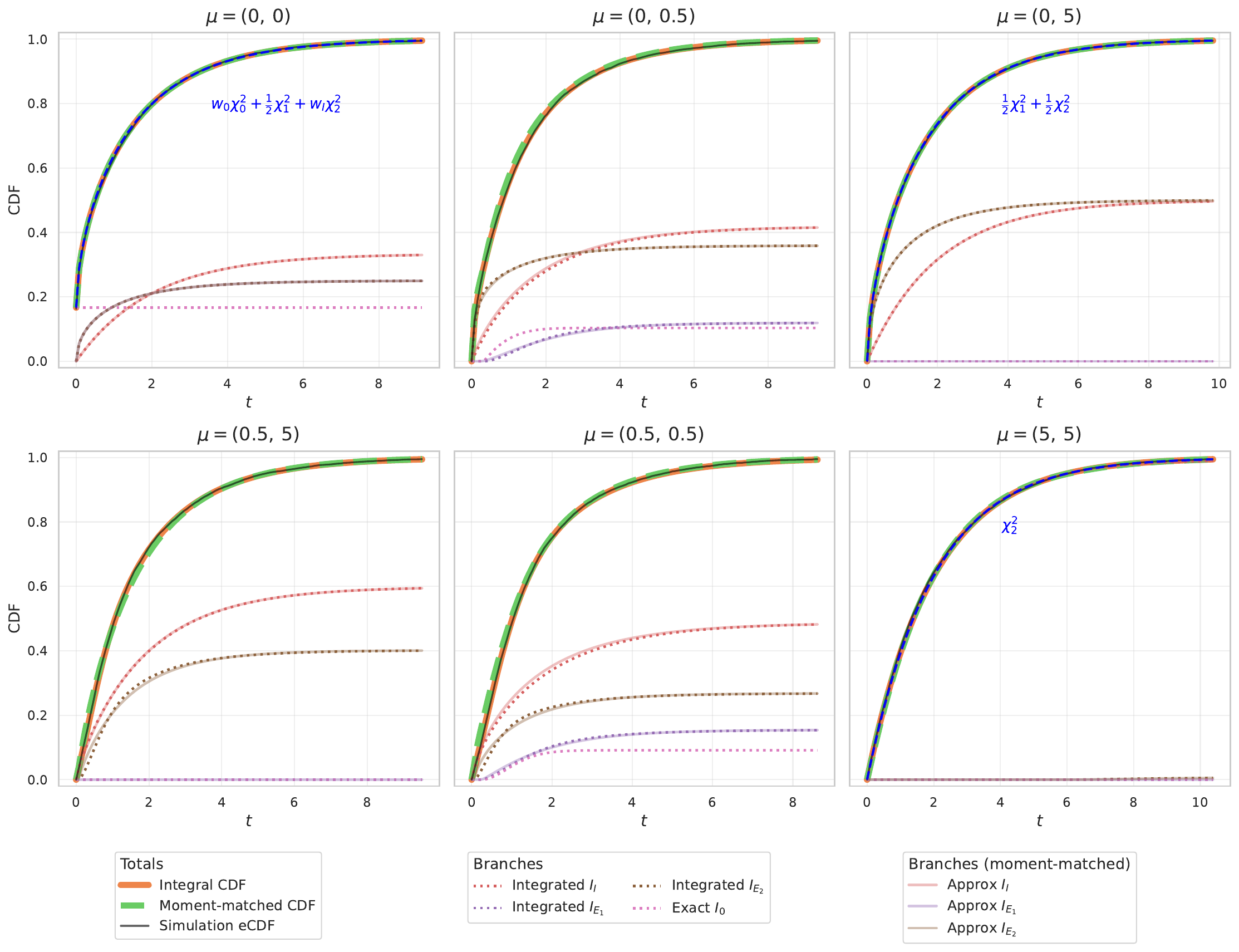}%
    }
    \caption{Null CDF of the likelihood ratio statistic for the two-bounded bivariate normal at six null values, arranged from the corner of the parameter space ($\mu=(0,0)$, top left) to its deep interior ($\mu=(5,5)$, bottom right). The variances are $\sigma_1^2=4$, $\sigma_2^2=1$ with correlation $\rho=0.5$, as in Figure~\ref{fig:2d-2params}. In each panel the total CDF is shown both by numerical integration of the branch contributions of Section~\ref{sec:2param_2bounded} (Integral CDF) and by the closed-form moment-matching approximation of Section~\ref{sec:moment-matching} (Moment-matched CDF), together with the empirical CDF of $10^5$ simulated statistics (Simulation eCDF) generated under the six different nulls. The four branch contributions are drawn separately: the interior $I_I$, the two edges $I_{E_1}$ and $I_{E_2}$, and the corner $I_0$.}
    \label{fig:CDFs-twoboundaries}
\end{figure}

\section{Numerical evaluation}\label{sec:numerical-evaluation}

\subsection{Accuracy of the null distribution}
\label{sec:coverage}
A confidence curve built from the likelihood ratio statistic is exact when its distribution under each null candidate coincides with the CDF used to calibrate it. Figure~\ref{fig:CDFs-twoboundaries} shows the CDF of the likelihood ratio statistic at six different null values for the bivariate normal model with both mean components bounded: from the corner of the parameter space to its deep interior. For each null candidate, we plot three estimates of the total CDF: numerical integration of the branch contributions from Section~\ref{sec:2param_2bounded} (Integral CDF), the moment-matching approximation of Section~\ref{sec:moment-matching} (Moment-matched CDF), and the empirical CDF of $10^5$ statistics simulated under the null (Simulation eCDF). We also display the four branch contributions $I_I$, $I_{E_1}$, $I_{E_2}$ and $I_0$ on their own. The confidence curves derived from either the Integral CDF or the Moment-matched CDF attain exact coverage if the Simulation eCDF overlaps them across the observed likelihood ratio statistic $q(\mu;x) = t$.

Placing the null on the boundary corner $\mu= (0,0)$, all four branches contribute. The corner term contributes with a constant term throughout $t$, corresponding to a point mass $w_0\chi^2_0$ with weight $w_0 = 0.25 - \arcsin(\rho)/(2\pi)$, while the two edges contribute equally with weight $0.25$ each, and their distributions are assumed to be $\chi^2_1$ on the respective branches. The interior carries the remaining weight $0.25 + \arcsin(\rho)/(2\pi)$ and contributes as a $\chi^2_2$ component, so the total CDF is the chi-bar-square distribution of \citet[Case 7]{SelfLiang1987}. 

In the intermediate cases, the branches contribute according to where the null is placed relative to the two boundary faces. For instance, in the upper middle panel ($\mu= (0, 0.5)$), the truth lies on the first boundary face but only a short standardised distance from the second. Here, the corner no longer gives a point mass, but a strictly increasing CDF and the two edges no longer overlap. Moving the null values further away from the second component's boundary, $\mu= (0, 5)$, the corner and second edge are no longer contributing, and the distribution is shared between the interior and the first edge with equal weights: this is precisely the distribution given in \citet[Case 6]{SelfLiang1987}, i.e.\ the asymptotic distribution of the likelihood ratio statistic when one of the two parameters of interest is on the boundary. Similarly, $\mu = (0.5, 5)$ reproduces the near-boundary transition of the one-bounded case in Section \ref{sec:2param_1bounded}, with only the interior and the first edge contributing. The panel $\mu = (0.5, 0.5)$ shows the regime where the null is placed close to both boundaries, where all four branches contribute.

As expected, only the interior branch is present in the deep-interior case: $\mu = (5,5)$. Here, the mean is sufficiently far from the boundary that the probability of $X$ landing in any of the other regions is negligible, and the total CDF coincides with the $\chi^2_2$ distribution, recovering Wilks' theorem. Thus, our boundary-aware limiting distribution of the likelihood ratio unifies a number of results present in the literature: Wilks' theorem and the mixtures of \citet[Cases 6 and 7]{SelfLiang1987} all stem as limits of a single distribution.

Across all six panels, the numerically integrated CDF agrees with the simulated empirical CDF to plotting accuracy, and the total closed-form moment-matching approximation is close to both. The largest discrepancy found between the moment-matching approximation and the numerically integrated CDFs is in the lower middle panel with the null parameter being $\mu = (0.5, 0.5)$. However, we deem this difference to still be small enough for the moment-matched approach to be considered a good approximation. In Section~\ref{sec:application}, we test their coverages for non-normal data.

\subsection{Confidence interval lengths and region areas}\label{sec:characteristics}

\begin{figure}
    \centering
    \includegraphics[width=0.75\linewidth]{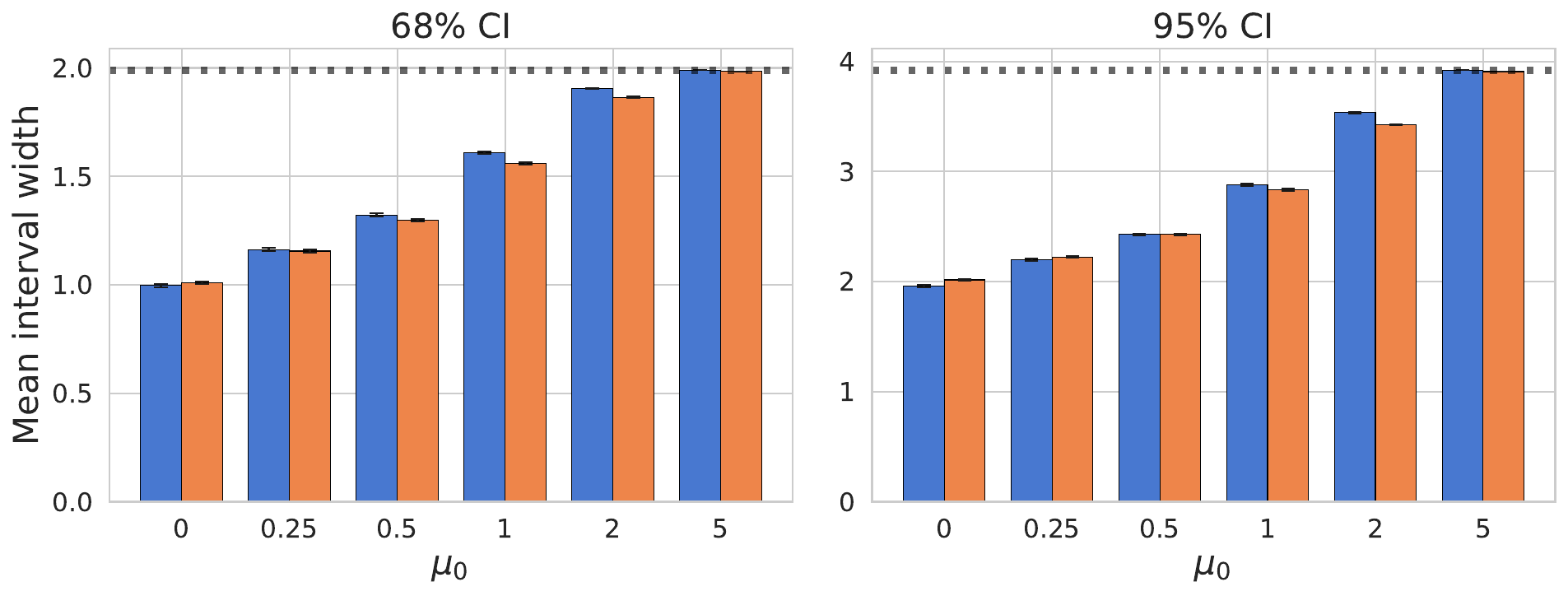}
    \includegraphics[width=0.75\linewidth]{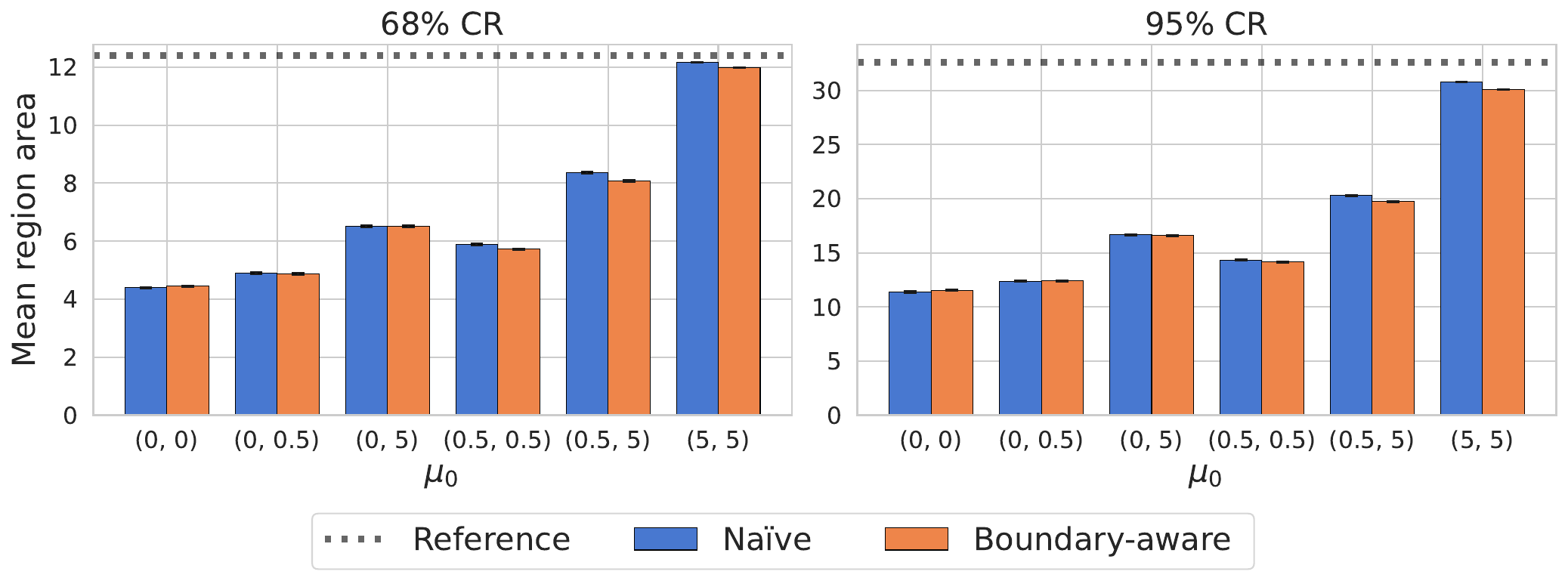}
    \caption{Mean widths (upper) and mean areas (lower) of $68\%$ and $95\%$ confidence intervals and regions for the bounded normal mean. Upper: univariate model $X \sim \mathcal{N}(\mu, 1)$, $\mu \geq 0$. Lower: bivariate model $X \sim \mathcal{N}_2(\mu, \Sigma)$, $\mu \in [0,\infty)^2$, with $\sigma_1 = 1$, $\sigma_2 = 2$, $\rho = 0.5$. The boundary-aware confidence curve (orange) is compared against the naïve confidence curve truncated to the parameter space (blue). Dotted lines mark the unbounded reference, based on $10^5$ MC simulations per true parameter setting $\mu_0$.}
    \label{fig:mean_widths_and_regions}
\end{figure}

For normal data with bounded mean components, both the naïve and the boundary-aware confidence curves attain nominal coverage, as shown for the scalar mean case in Figure~\ref{fig:cc_constrained_example}. When two constructions both attain correct coverage, they can be compared on two further criteria: the bias of the associated point estimator, and the width of the confidence intervals or the area of the confidence regions, with narrower intervals or regions preferred; see, for instance, \citet[Chapter 5]{SchwederHjort2016} and \citet[Section 5.3]{CunenHjort2022}.

The biases of the two estimators (proper and naïve MLE) are evaluated in Appendix~\ref{app:Bias}. The naïve MLE is both mean and median unbiased for normally distributed data, but only because it is free to take values outside the parameter space. The boundary-aware approach gives up mean unbiasedness (and median too, in the two-parameter case) in exchange for an estimator which always lies in the parameter space and does not return empty intervals, as discussed in \citet{Mandelkern2002}.

Figure \ref{fig:mean_widths_and_regions} shows the mean widths and mean areas of the $68\%$ and $95\%$ confidence intervals and regions as a function of the true value $\mu_0$ obtained via simulation. In both panels, the boundary-aware confidence curve is compared with the curve obtained from the naïve estimator via the chi-square construction in Eq.~\eqref{eq:cc_LR}, truncated to the parameter space post hoc.

The upper panel shows the mean width of the $68\%$ and $95\%$ confidence intervals for the bounded mean of a univariate normal model with known variance (Section~\ref{sec:1d-WuNeale}). For true means on or very close to the boundary, $\mu_0 = 0$ and $\mu_0 = 0.25$, the boundary-aware intervals are slightly wider than the confidence intervals from the naïve approach. This should not be read as an advantage of the naïve approach. It truncates the part of each interval that lies below zero, and the closer $\mu_0$ is to the boundary, the more often a large part, or all, of the interval is removed. The boundary-aware intervals, in contrast, are never empty at confidence levels of $50\%$ or higher, making them slightly wider here. At intermediate distances from the boundary ($\mu_0 = 0.5, 1, 2$), however, the boundary-aware intervals are consistently shorter. At $\mu_0 = 5$ the boundary constraint is effectively inactive, and both methods produce intervals of equal width, converging to the unbounded reference $2z_{1-\alpha/2}\sigma$, where $z_{1-\alpha/2} = \Phi^{-1}(1-\alpha/2)$ is the standard normal $1-\alpha/2$ quantile.

The lower panel shows the mean area of the $68\%$ and $95\%$ confidence regions for a bounded mean vector from a bivariate normal model with known variances and correlation $\rho = 0.5$ (Section~\ref{sec:2param_2bounded}). The mean areas of the $68\%$ and $95\%$ confidence regions are plotted for six null values ranging from the corner $\mu_0 = (0,0)$ to the deep interior $\mu_0 = (5,5)$, as in Figure~\ref{fig:CDFs-twoboundaries}. The pattern is similar: the boundary-aware confidence regions are smaller than or comparable to the truncated naïve approach regions across all null values, with the largest difference at intermediate boundary distances such as $\mu_0 = (0.5, 5)$. At $\mu_0 = (5,5)$ both methods converge to the unbounded ellipse area $\pi\sqrt{|\Sigma|} F_{\chi^2_2}^{-1}(1-\alpha)$, while the boundary-aware approach has slightly lower area even here. 

\section{Application: The Signal-Plus-Background Experiment}\label{sec:application}

The signal-plus-background experiment is a standard analysis in high-energy physics, arising, for instance, in the analysis of particle collision data collected by experiments at the Large Hadron Collider at CERN. The goal is to evaluate how much new physics the data support, typically by constructing confidence intervals on signal rates; see, for instance, \citet{cowan_asymptotic_2011} and \citet{Algeri2020}. Since a confidence distribution collects all confidence intervals into a single object, it provides a natural summary of such inference, a point also made by \cite{FraserReidWong2004}. In this paper, we encourage reporting results in the form of confidence curves, as they provide a convenient representation and visualisation of uncertainty for multidimensional parameters. In the univariate setting, they can easily be transformed into the one-sided standard CD form using Eq.~\eqref{eq:CDtoCC-transformation}.
\subsection{Setup}

Consider a detector recording events over a spectrum divided into $d$ bins, with the number of events in each bin $i = 1,\ldots,d$ given by
\begin{equation}
    N_i \sim \operatorname{Poisson}(\mu_{i} + \beta_{i}),
    \label{eq:signal_plus_background}
\end{equation}
where $\mu = (\mu_{1}, \ldots, \mu_{d}) \geq 0$ is the signal rate, the parameter of interest,  and $\beta = (\beta_{1}, \ldots, \beta_{d}) > 0$ the background rate, a nuisance parameter, with unknown true values $\mu_0$ and $\beta_0$, respectively. 

The background rate $\beta$ is often not known a priori and must be estimated from auxiliary data. We model this as
\begin{equation}
    X_{\beta} \sim \mathcal{N}_d(\beta, \Sigma_\beta),
    \label{eq:b-auxiliary}
\end{equation}
where $\Sigma_\beta$ is a known covariance. The full collection of independent data are $(N_1, \ldots, N_d, X_\beta)$ and the profile likelihood for $\mu$ is obtained by maximising the joint log-likelihood over $\beta$ at each fixed $\mu$. We take the true background to be sufficiently large compared to its uncertainty, so that boundary effects on $\beta$ are negligible. This is typical in collider searches, where expected backgrounds commonly range from tens to many thousands of events per bin \citep[e.g.,][]{aaboud_search_2017, aad_search_2023, aad_search_2024}.

When handling a similar example, \citet{FraserReidWong2004} impose only a boundary for $\mu + \beta > 0$, required for the Poisson rate to define a valid distribution, and truncate the resulting confidence intervals in $\mu$ to its parameter space.\footnote{The authors use a third-order approximation to their confidence distribution, based on the modified likelihood root to improve the finite sample accuracy in the presence of an estimated background. In our examples, the true background is chosen large enough so that the first-order approximation suffices, so no such correction is applied.} Their treatment of the boundary is thus the one we have termed the \emph{naïve} approach, and we refer to the resulting confidence curves as \emph{naïve CCs}. We compare this approach with our boundary-aware approach, which uses the proper MLE. In this case, $\mu \geq 0$ is enforced directly, so that the boundary corrections of the preceding sections apply when the null value of $\mu$ lies near zero. The confidence curves constructed using this approach are referred to as \emph{boundary-aware CCs}. 

\begin{figure}[htbp]
    \centering
    \makebox[\textwidth][c]{%
        \includegraphics[width=1.5\textwidth]{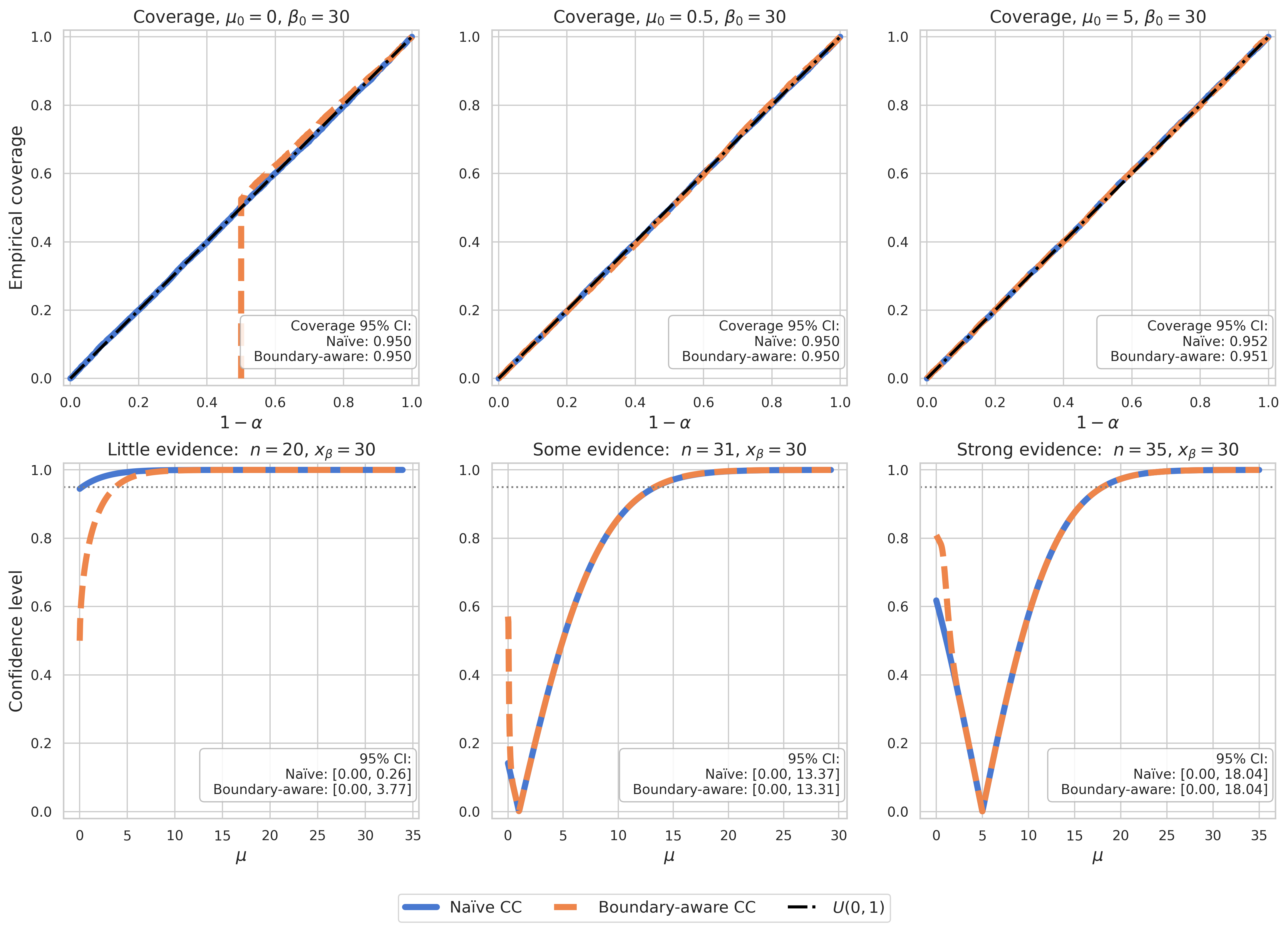}%
    }
    \caption{Simulated coverages (upper) and confidence curves from three different observations (lower) of the signal-plus-background experiment with one signal parameter. True signal values $\mu_0 \in \{0, 0.5, 5\}$, background $\beta_0 = 30$, auxiliary variance $\sigma_\beta^2 = 1$. The boundary-aware CC is compared against the naïve CC. Coverage is tested by the empirical CDF of $\mathrm{cc}(\mu_0; N, X_\beta)$ under repeated sampling, using $10^5$ MC simulations.}
    \label{fig:cc-poisson-1d}
\end{figure}

The distributions found in Section \ref{sec:dist} were derived analytically for normal data with known covariance. However, they can be extended to any regular parametric model in the usual asymptotic sense \citep[Chapter 16]{vaart_asymptotic_1998}. For the signal-plus-background model, the normality is discussed in detail in \citet[Section 3.1]{cowan_asymptotic_2011}, who use the Wald expansion to write the profile likelihood ratio as a quadratic form in the signal parameter. To first order, the inverse profile information $\Sigma_\mu$, evaluated at the profiled background $\hat\beta(\mu)$, replaces $\Sigma$. The MLE $\hat\mu$ is the projection of $\tilde\mu$ onto $[0,\infty)^d$ in the $\Sigma_\mu^{-1}$-metric, and the corrections derived in Section \ref{sec:dist} apply to the likelihood ratio statistic built from it.

We treat two cases: $d = 1$ in Section~\ref{sec:app-1d}, where Section~\ref{sec:1d-WuNeale} applies, and $d = 2$ in Section~\ref{sec:app-2d}, where we use both the integral form of Section~\ref{sec:2param_2bounded} and the moment-matched approximation of Section~\ref{sec:calibration}. In both cases, we set the true backgrounds $\beta_{i,0} = 30$, which is large enough for the counts $N_i$ to have an approximate normal distribution, and choose small values of the true signal. The normal approximation underlying Section~\ref{sec:dist} therefore holds, while the standardised distance to the boundary remains at the intermediate values for which our corrections are needed.

\subsection{One parameter of interest}\label{sec:app-1d}

With a single channel, observing $N = n$ and $X_\beta = x_\beta$, the joint log-likelihood is 
$$\ell(\mu, \beta) \propto n \log(\mu + \beta) - (\mu + \beta) - (x_\beta - \beta)^2 / (2\sigma_\beta^2).$$ 
In order to construct a confidence curve in $\mu$ alone, we must first profile the likelihood over $\beta$. Furthermore, to apply the corrections in Section~\ref{sec:1d-WuNeale}, we need the standardised boundary distance $\delta = \mu/\sigma_\mu$, where $\sigma_\mu^2$ is the variance of the MLE $\hat{\mu}$. We approximate this using the inverse observed information. The full observed information matrix for $(\mu, \beta)$ is

\setlength{\arraycolsep}{10pt}
$$
  \mathcal{I}(\mu, \beta) = \begin{pmatrix} n(\mu+\beta)^{-2} & n(\mu+\beta)^{-2} \\ n(\mu+\beta)^{-2} & n(\mu+\beta)^{-2} + \sigma_\beta^{-2} \end{pmatrix},
$$
and the profile information for $\mu$ is the Schur complement $\mathcal{I}_{\mu\mu} - \mathcal{I}_{\mu\beta}^2 / \mathcal{I}_{\beta\beta}$, whose inverse gives $\sigma_\mu^2 = (\mu+\beta)^2/n + \sigma_\beta^2$. We evaluate $\sigma_\mu^2$ at the profiled nuisance $\hat{\beta}(\mu)$. 

\begin{figure}[htbp]
    \centering
    \makebox[\textwidth][c]{%
        \includegraphics[width=1.5\textwidth]{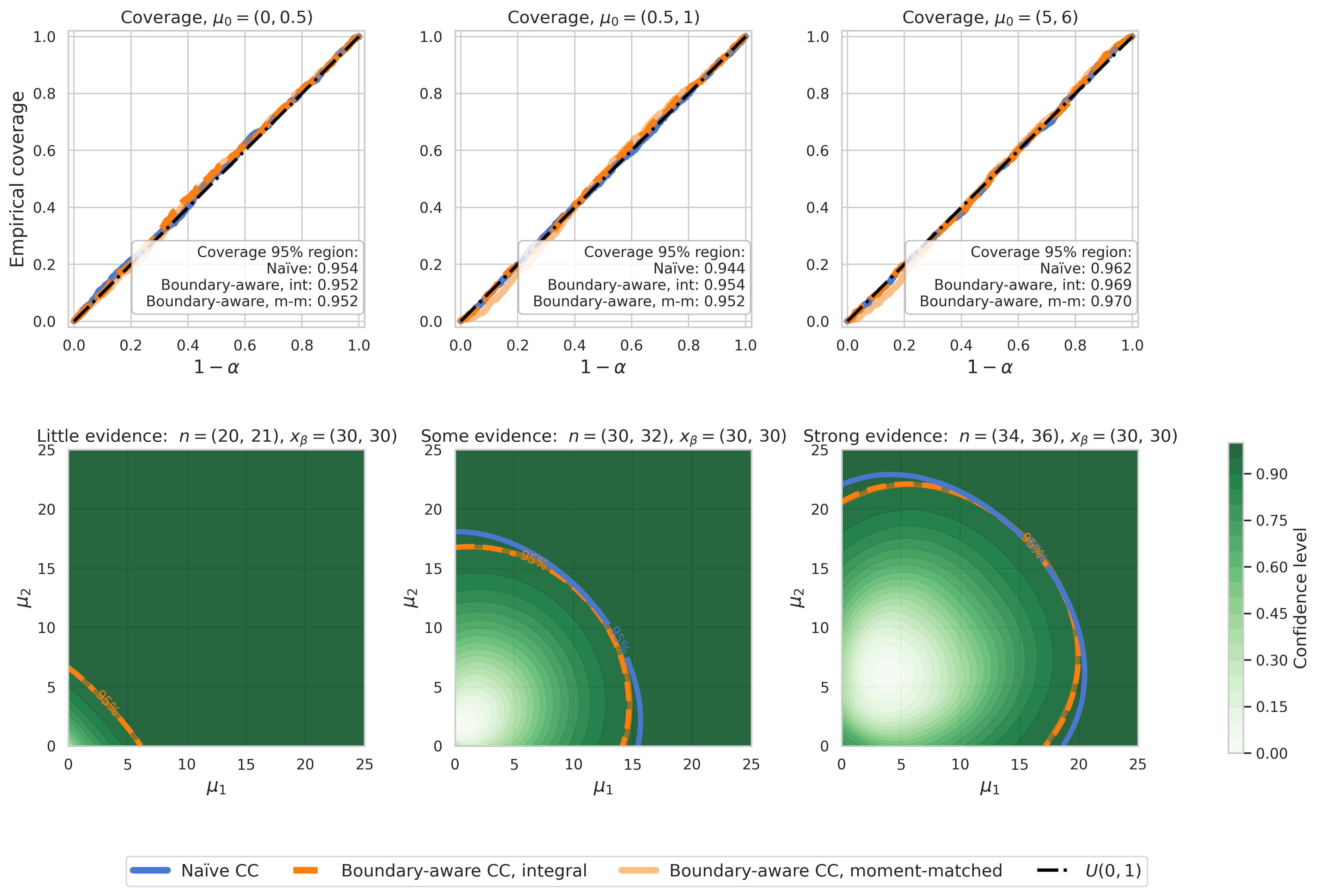}%
    }
    \caption{Simulated coverages and bivariate confidence regions for the signal-plus-background experiment with two signal parameters. True signal values $\mu_0 \in \{(0,0.5),\, (0.5, 1),\, (5,6)\}$, background $\beta_0 = (30,30)$, auxiliary known covariance $\Sigma_\beta$ with $\sigma_{\beta_1} = \sigma_{\beta_2} = 1$ and $\rho = 0.5$. Upper panels: empirical CDF of the confidence curve evaluated at the true $\mu_0$, based on $10^5$ MC simulations. Lower panels: the full bivariate confidence curve from the boundary-aware CC along with overlays of $95\%$ boundary-aware confidence region in orange (dashed: exact integral-based construction, solid: moment-matched approximation) and truncated naïve approach confidence region in blue for three different observations of the data $(N_1, N_2, X_\beta)$.}
    \label{fig:cc-poisson-2d}
\end{figure}

Figure~\ref{fig:cc-poisson-1d} shows confidence curves and simulated coverage of all levels $1-\alpha$ with underlying true background $\beta_0 = 30$ and $\sigma_\beta = 1$. The upper panels display example confidence curves from three observations of the pair $(N, X_\beta) = (n, x_\beta)$, representing increasing levels of evidence for a non-zero signal. In the left panel the observed count is well below the estimated background ($n = 20$, $x_\beta = 30$), so the data provide very little evidence for a signal. The boundary-aware CC equals $0.5$ at $\mu = 0$, the smallest value it can take at the boundary, because the chi-bar-square null distribution there has a point mass of $0.5$ at zero. The naïve CC, by contrast, gives an empty interval for all confidence levels below $94\%$. This is an instance of the empty interval problem discussed by \citet{Mandelkern2002}. At $95\%$ the naïve interval is $[0, 0.26]$, while the boundary-aware interval is $[0, 3.77]$. The middle panel shows an observation where $n$ slightly exceeds $x_\beta$ ($n = 31$, $x_\beta = 30$), suggesting a small signal. Here the two curves are similar, though the boundary-aware CC takes a higher value at $\mu = 0$. In the right panel, there is stronger evidence for a non-zero signal ($n = 35$, $x_\beta = 30$), and the two curves agree even more closely, though again the boundary-aware CC takes a higher value at $\mu = 0$.

The lower panels show simulated coverage of the CCs at the truths $\mu_0 = 0$, $0.5$ and $5$, based on $10^5$ MC simulations. For $\mu_0 = 0$, the boundary-aware CC is slightly overcovering for confidence levels between 50\% and 90\%. However, in all three panels, both methods produce close-to exact coverage at 95\% confidence level. 

\subsection{Two parameters of interest}\label{sec:app-2d}
With two channels, $i = 1, 2$, the data are the independent counts $N_1, N_2$ from Eq.~\eqref{eq:signal_plus_background} and the bivariate auxiliary measurement $X_\beta$ from Eq.~\eqref{eq:b-auxiliary}. We profile over $\beta$ at each fixed $\mu \geq 0$ and, as in the one-dimensional case, approximate the covariance of $\hat{\mu}$ by the inverse Schur complement of the observed information,
\begin{equation*}
\Sigma_\mu = \operatorname{diag}\left\{(\mu_1+\beta_1)^2/n_1,\, (\mu_2+\beta_2)^2/n_2\right\} + \Sigma_\beta,
\end{equation*}
evaluated at $\hat{\beta}(\mu)$. The boundary-aware CC is then calibrated with this covariance using both the integral representation of Section \ref{sec:bivar-formulation} and the moment-matched approximation of Section \ref{sec:mm2}.

Figure~\ref{fig:cc-poisson-2d} shows simulated coverages and bivariate confidence curves with true backgrounds $\beta_0 = (30, 30)$ and $\Sigma_\beta$ composed of $\sigma_{\beta_1} = \sigma_{\beta_2} = 1$ and $\rho = 0.5$. There are three settings for the true signal parameters in the coverage plot: $\mu_0 = (0,0.5)$, $(0.5, 1)$ and $(5, 6)$, which are based on $10^5$ MC repetitions. The moment-matched approximation deviates slightly from nominal coverage at low confidence levels, most visibly in the middle and right upper plots, while the integral-based curve stays close to the diagonal throughout. At the 95\% level, all three curves have close-to-nominal coverage.

The lower panels display the full bivariate boundary-aware CC, with the boundary-aware and naïve $95\%$ confidence region overlaid for three observations. Similar to the lower panel of Figure~\ref{fig:cc-poisson-1d}, we show three settings with increasing levels of evidence. 
In the left panel the counts fall far below the estimated background ($n = (20, 21)$, $x_\beta = (30, 30)$), giving very little evidence for a non-zero signal. Here, the boundary-aware 95\% region is narrow, while the naïve 95\% region is empty, and thus not showing up in the plot. 
In the middle panel, the second count slightly exceeds the background ($n = (30, 32)$, $x_\beta = (30, 30)$), and the two regions are closer but still distinguishable. In this setting, the 95\% confidence region from boundary-aware CC is narrower than the one from the naïve CC. In the right panel, the counts clearly exceed the background ($n = (34, 36)$, $x_\beta = (30, 30)$), and again we find that the 95\% confidence region is narrower for the boundary-aware CC than for the naïve one.

\section{Discussion}\label{sec:discussion}
In this paper we showed how to construct confidence curves for one or two bounded parameters from the likelihood ratio statistic. Its limiting distribution used to construct the confidence curves contains Wilks' theorem and the chi-bar-square distributions of \citet[Cases 6 and 7]{SelfLiang1987} as limits, and covers the near-boundary regime between them. We gave this distribution in two forms: an integral form, exact for normal data with known covariance up to numerical integration precision, and a moment-matched approximation. The two closely agree in our simulations, while the latter is arguably easier for practitioners to implement.

The resulting confidence curves achieve correct asymptotic coverage and are often narrower than those built on the naïve approach, in which the estimator is allowed to take values outside the parameter space. The exception is when the true parameter values sit on or very close to their boundaries, where the corrected intervals are slightly wider; following \citet[Chapter 5]{SchwederHjort2016}, we therefore gain power using the boundary-aware approach everywhere else.

Moreover, the boundary-aware approach has the practical advantage of never returning an empty interval. At typical confidence levels, the naïve route can return empty intervals for unlucky observations, i.e.,\ plausible observations for which the naïve estimator falls far outside the parameter space. This phenomenon was highlighted by \citet{Mandelkern2002} and occurs in our simulations (see the lower left panels of Figures~\ref{fig:cc-poisson-1d} and~\ref{fig:cc-poisson-2d}). These empty intervals enter the average with value zero (or slightly above zero) and so deflate the mean naïve confidence lengths and areas. The boundary-aware construction never returns an empty interval, which matters in practice: an expensive experiment cannot simply be rerun when its analysis returns an empty interval. Thus, since unlucky observations may happen, we argue that this outcome is undesirable.

As a practical example, we applied the boundary-aware approach to the signal-plus-background experiment. Here, everything behaved as intended, and the moment-matched approximation of Section~\ref{sec:calibration} works well with small discrepancies at most. Several physics analyses of this type already adopt the boundary-aware approach, as discussed in Section \ref{sec:intro}. However, they calibrate it against a simulated distribution rather than an asymptotic one. In practice, these simulations require massive computing power \citep{agapopoulou_critical_2025}. Our results supply the asymptotic distribution directly, removing the need to simulate at every point of the parameter space. Thus, our results give an accurate and computationally cheaper route for constructing confidence intervals, regions, and the confidence curve as a whole.

We have deliberately stopped at two dimensions. Two bounded parameters are common in practice as a jointly reported pair, with the remaining parameters profiled out. A bivariate  confidence region  can be drawn and read, whereas regions in three or more dimensions cannot be displayed in the same way. The parameters outside the reported pair are then treated as nuisances. One issue we have not addressed is the correction needed when such nuisance parameters are themselves bounded. This has been studied for testing on the boundary \citep{SelfLiang1987, kopylev_asymptotic_2011}, and extending it to confidence curves would be a natural next step.

\section*{Acknowledgements}
We thank Anders Kvellestad for discussions and contributions to this paper, in particular on the statistical methodology used in high-energy physics. We also thank Are Raklev, both for valuable inputs on this work and for arranging the travel grant for HBR. 

HBR, CBS and RDB were supported by the Research Council of Norway (RCN) through the FRIPRO project PLUMBIN’ (project no. 323985) and the Centre of Excellence INTEGREAT (project no. 332645). HBR was also in part funded by an RCN mobility grant for a research stay at the University of Padova (project no. 361953). ARB was supported by the Italian Ministry of University and Research (MUR) under the PRIN 2022 call (final rankings scrolling), project “Latent variable models and dimensionality reduction methods for complex data” (project no. 20224CRB9E, CUP C53C24000730006).
\bibliographystyle{apalike}
\bibliography{references} 
\newpage

\appendix

\section{Bias of the MLE in the normal model}\label{app:Bias}

\subsection{One-dimensional parameter of interest}\label{app:bias-1dim}
For the univariate normal model with one observation, the naïve MLE is $\tilde{\mu}(X) = X$ and is mean unbiased, while the MLE for $M_1 = [0,\infty)$ is $\hat{\mu}(X) = \max\{0, X\}$. The truncation lifts every negative value of $X$ up to zero and leaves the rest unchanged, yielding the mean bias of the MLE
\begin{equation*}
  E[\hat{\mu}(X)] - \mu_0 =\sigma\phi(\mu_0/\sigma) - \mu_0\Phi(-\mu_0/\sigma) > 0 \quad \forall \mu_0 \ge 0.
\end{equation*}
The mean bias is largest on the boundary, where it takes the value $ \sigma/\sqrt{2\pi}$. By $\mu_0 = 2\sigma$, only about $2.3\%$ of the distribution lies below zero, and the bias is below 
$0.01\sigma$.

The median behaves differently. Truncation only moves the mass below zero, and since $\mu_0 \ge 0$ this is at most half of the distribution, and the median does not move, i.e., $\operatorname{med}\{\hat{\mu}(X)\} = \operatorname{med}(X) = \mu_0$. Both the naïve MLE and the MLE are therefore median unbiased on all of $M_1$. For non-normal data, neither estimator is in general mean or median unbiased.

\subsection{Two-dimensional parameter of interest}\label{app:bias-2dim}
Let $X \sim \mathcal{N}_2(\mu,\Sigma)$ with the parameter space of the mean $M_2 = [0,\infty) \times [0,\infty)$. The naïve MLE is again $\tilde{\mu}(X) = X$ and is both mean and median unbiased, while the boundary-aware MLE is generally neither. Outside limiting cases, the mean and median bias have no simple mathematical expression. Thus, we rather report the MC simulated mean and median biases for unit variances with correlations $\rho = -0.5, 0$ and $+0.5$ over a range of true values $\mu_0$ in Table~\ref{tab:bivariate-bias}.

The $\rho = 0$ block reproduces the one-dimensional result as the individual estimators $\hat{\mu}_1, \hat{\mu}_2$ are independent, i.e.\ one component carries no information about the other, so the problem separates.  For $\rho \neq 0$ this is no longer the case. Even when one mean component lies far from its own boundary, so that its constraint never binds, it is still shifted whenever the constraint on the other component becomes active. The rows $\mu_0 = (0,5)$ and $(0.5,5)$ confirm this: the mean biases for the second mean component are $+0.200$ and $+0.099$ for $\rho = +0.5$, the same magnitudes with reversed sign for $\rho = -0.5$, while for the first component they are unchanged at $0.399$ and $0.198$, respectively. Moreover, at $\mu_0 = (0,0)$ the mean bias is $0.472$ for $\rho = +0.5$, against $0.399$ under independence and $0.273$ for $\rho = -0.5$, and the same ordering is seen at $\mu_0 = (0.5,0.5)$. When both components are far from their boundaries, as in the case $\mu_0 = (5,5)$, both mean and median bias are zero to simulation precision for all three correlations.

Median-unbiasedness does not survive in the two-dimensional case for $\rho \neq 0$, as seen in Table~\ref{tab:bivariate-bias}. The absolute value of median bias stays below $0.192$, against the largest mean bias of $0.472$. Both vanish once $\mu_1$ and $\mu_2$ are far from the boundary. 

\begin{table}[t]
\centering
\caption{Simulated mean and median bias of the two-dimensional MLE in the bivariate normal model with both mean parameters bounded, $M_2=[0,\infty)\times[0,\infty)$. The naïve MLE is both mean and median unbiased and therefore is omitted. Unit variances are used, so $\delta=\mu_1$ and $\delta' = \mu_2$. The results are based on $10^8$ MC repetitions.}
\label{tab:bivariate-bias}
\vspace{0.5em}
\makebox[\textwidth][c]{%
\begin{tabular}{llcc}
\hline
\rule{0pt}{2.5ex}$\rho$ & $\mu_0$ & Mean bias & Median bias \\[0.3em]
\hline
\rule{0pt}{2.5ex}$-0.5$
 & $(0,\,0)$     & $(+0.273,\,+0.272)$ & $(+0.000,\,+0.000)$ \\[0.2em]
 & $(0,\,0.5)$   & $(+0.328,\,+0.035)$ & $(+0.000,\,-0.191)$ \\[0.2em]
 & $(0,\,5)$     & $(+0.399,\,-0.199)$ & $(+0.000,\,-0.191)$ \\[0.2em]
 & $(0.5,\,0.5)$ & $(+0.112,\,+0.111)$ & $(-0.088,\,-0.089)$ \\[0.2em]
 & $(0.5,\,5)$   & $(+0.198,\,-0.099)$ & $(-0.000,\,-0.088)$ \\[0.2em]
 & $(5,\,5)$     & $(-0.001,\,-0.000)$ & $(-0.000,\,+0.000)$ \\[0.4em]
\hline
\rule{0pt}{2.5ex}$+0.0$
 & $(0,\,0)$     & $(+0.399,\,+0.399)$ & $(+0.000,\,+0.000)$ \\[0.2em]
 & $(0,\,0.5)$   & $(+0.399,\,+0.198)$ & $(+0.000,\,-0.000)$ \\[0.2em]
 & $(0,\,5)$     & $(+0.399,\,+0.000)$ & $(+0.000,\,+0.000)$ \\[0.2em]
 & $(0.5,\,0.5)$ & $(+0.198,\,+0.198)$ & $(+0.000,\,+0.000)$ \\[0.2em]
 & $(0.5,\,5)$   & $(+0.198,\,-0.000)$ & $(+0.000,\,-0.000)$ \\[0.2em]
 & $(5,\,5)$     & $(-0.000,\,-0.000)$ & $(+0.000,\,+0.000)$ \\[0.4em]
\hline
\rule{0pt}{2.5ex}$+0.5$
 & $(0,\,0)$     & $(+0.472,\,+0.472)$ & $(+0.191,\,+0.191)$ \\[0.2em]
 & $(0,\,0.5)$   & $(+0.427,\,+0.314)$ & $(+0.089,\,+0.191)$ \\[0.2em]
 & $(0,\,5)$     & $(+0.399,\,+0.200)$ & $(+0.000,\,+0.191)$ \\[0.2em]
 & $(0.5,\,0.5)$ & $(+0.247,\,+0.247)$ & $(+0.089,\,+0.089)$ \\[0.2em]
 & $(0.5,\,5)$   & $(+0.198,\,+0.099)$ & $(+0.000,\,+0.089)$ \\[0.2em]
 & $(5,\,5)$     & $(-0.000,\,+0.000)$ & $(-0.000,\,-0.000)$ \\[0.4em]
\hline
\end{tabular}%
}
\end{table}

\end{document}